\documentclass[aps,prd,preprintnumbers,superscriptaddress,groupedaddress,nofootinbib]{revtex4}

\pdfoutput=1

\usepackage{multirow}
\usepackage{subfigure}
\usepackage{graphicx}
\usepackage{amsfonts}
\usepackage{amsmath,amssymb}
\usepackage{slashed}
\usepackage{feynmp}
\usepackage{caption}
\usepackage{url}
\usepackage{color}
\usepackage{cancel}

\begin{document}

\title{Scalar Simplified Models for Dark Matter}

\author{Matthew R.~Buckley}
\affiliation{Department of Physics and Astronomy, Rutgers University, Piscataway, NJ 08854, USA}
\author{David Feld}
\affiliation{Department of Physics and Astronomy, Rutgers University, Piscataway, NJ 08854, USA}
\author{Dorival Gon\c{c}alves}
\affiliation{Institute for Particle Physics Phenomenology, Department of Physics, Durham University, United Kingdom}

\preprint{IPPP/14/92,  DCPT/14/184}
\date{\today}

\begin{abstract}
We introduce a set of minimal simplified models for dark matter interactions with the Standard Model, connecting the two sectors via either a scalar or pseudoscalar particle. These models have a wider regime of validity for dark matter searches at the LHC than the effective field theory approach, while still allowing straightforward comparison to results from non-collider dark matter detection experiments. Such models also motivate dark matter searches in multiple correlated channels. In this paper, we constrain scalar and pseudoscalar simplified models with direct and indirect detection experiments, as well as from existing LHC searches with missing energy plus tops, bottoms, or jets, using the exact loop-induced coupling with gluons. This calculation significantly affects key differential cross sections at the LHC, and must be properly included. We make connections with the Higgs sector, and conclude with a discussion of future searches at the LHC.
\end{abstract}
\maketitle

\section{Introduction \label{sec:intro}}

The case for the existence of dark matter is strong. Decades of evidence from multiple independent lines \cite{Zwicky:1933gu,Rubin:1980zd,Olive:2003iq,Ade:2013zuv} reveal that this form of matter has a significant role in the composition and evolution of our Universe (for a review, see {\it e.g.}, Ref.~\cite{Bertone:2004pz}). No particle in the Standard Model is a suitable candidate for dark matter and so we need new physics to explain it. Though we lack evidence of the nature of the dark sector, if particle dark matter has a mass at the TeV scale or lower and was ever in thermal equilibrium in the early Universe, we have good reason to expect interactions with the visible sector to be within reach of our present experiments. However, this is of course not guaranteed. 

Perhaps the best known example of such dark matter is a weakly-interacting massive particle which becomes a thermal relic with the appropriate energy density after freeze-out. This type of dark matter is realized in many extensions of the Standard Model introduced to solve other problems of a theoretical nature ({\it e.g.}~Naturalness and Hierarchy). However, looking beyond this class of dark matter, even models of non-thermal dark matter often require significant annihilation cross sections into either the Standard Model or some hidden sector, so as not to overclose the Universe \cite{Buckley:2011kk}. It is therefore well-motivated to search for dark sector particles in a range of experiments, including the Large Hadron Collider (LHC).

When looking for dark matter, we can cast the experimental reach in terms of specific models of dark matter which are UV-complete. These models usually have a number of additional new particles with more significant interactions with the Standard Model than the dark matter itself. The canonical example of this sort is the supersymmetric neutralino, which is accompanied by a host of new charged and colored superpartners.  Despite the advantage of UV-complete models, interpreting results in this way has some drawbacks: $i$) the results may be difficult to recast for new models; $ii$) correlating results with non-collider experiments may be very dependent on UV-complete parameters; $iii$) focusing on a specific high-energy model runs the risk of overlooking other experimentally interesting channels; and $iv$)  tuning the experimental selection criteria could reduce the sensitivity to other types of dark matter.

In order to approach the problem in a somewhat model-independent way while still allowing for comparison between different classes of experiments, it has been useful to present the results of experimental searches in an effective field theory (EFT) framework \cite{Cao:2009uw,Goodman:2010yf,Goodman:2010ku}. The EFT approach assumes contact term interactions between dark matter and SM particles with the particle(s) connecting the two sectors integrated out of the low-energy spectrum. The validity of the EFT approach diminishes in the regime where the momentum transfer cannot be neglected relative to the (unknown) mass of the heavy particles. For direct detection this condition is usually satisfied, as long as mediators are not extremely light, as the momentum scale is on the order of 10 keV. Indirect detection and thermal freeze-out involve the annihilation of non-relativistic dark matter and so the EFT is applicable as long as the mediator is significantly heavier than twice the dark matter mass, assuming no additional new particles in the theory \cite{Abdallah:2014hon}. 

However, when considering the production of dark matter at particle colliders through high $p_T$ visible particles recoiling against invisible dark matter~\cite{Birkedal:2004xn,Feng:2005gj,Beltran:2008xg,Konar:2009ae,Beltran:2010ww,Bai:2010hh,Rajaraman:2011wf,Fox:2011pm,Bai:2012xg,Fox:2012ee,Carpenter:2012rg}, the momentum transfer in dark matter pair production events is large enough to render the EFT assumption invalid for a significant range of dark matter masses, couplings, and mediator masses \cite{Bai:2010hh,Fox:2011fx,Fox:2011pm,Shoemaker:2011vi,Fox:2012ee,Weiner:2012cb,Busoni:2013lha,Buchmueller:2013dya,Buchmueller:2014yoa,Busoni:2014sya,Busoni:2014haa}. As the momentum flowing through the production diagram is proportional to both the transverse momentum of the dark matter particles ({\it i.e.}~the missing transverse momentum, or MET) and the transverse momentum of recoiling visible particles required for the trigger, this issue will  be even more pressing at the LHC Run-II, as the trigger requirements on MET and jet $p_T$ will be higher than those used in Run-I.  Rather than viewing the invalidity of the EFT formalism as a drawback, it should be seen as an optimistic statement: if dark matter is being produced at colliders, it is generally the case that new mediating particles are being produced as well. 

As we look to interpret results from dark matter experiments and design new search strategies at the LHC, a balance should be struck between the very general (but often inapplicable) EFT approach and a full theory like supersymmetry. One solution has been found in {\it Simplified Models} \cite{Alwall:2008ag,Alves:2011wf,Goodman:2011jq}, which resolve the contact interaction into a single exchange particle, without adding in the full complexity of a UV-complete model. By specifying the spin and gauge quantum numbers of the dark matter and the mediators, the parameter space can be made relatively small, allowing an easy conversion of bounds between experiments and theories. Previous papers
have discussed colored mediators \cite{An:2013xka,DiFranzo:2013vra,Papucci:2014iwa}, which result in $t$-channel production of dark matter in a manner very similar to squarks in supersymmetry. Other works have considered vector and axial vector $Z'$ models~\cite{An:2012va,Frandsen:2012rk,Busoni:2014haa}, which cause $s$-channel dark matter production at colliders.

In this paper we consider a class of simplified models with a spin-0 scalar or pseudoscalar mediator, which allows $s$-channel production of dark matter from Standard Model partons at the LHC. These models are attractive in their simplicity, requiring only a minimal extension of the Standard Model's particle content. New scalars or pseudoscalars can also be easily accommodated in extended Higgs sectors, and it is not unreasonable to expect the Higgs to have contact with the dark sector. 
As with other simplified models, scalar mediators predict LHC signatures in a number of correlated channels; this can be used to our advantage when designing new searches. 

\begin{figure}[h!]
\includegraphics[width=0.73\columnwidth]{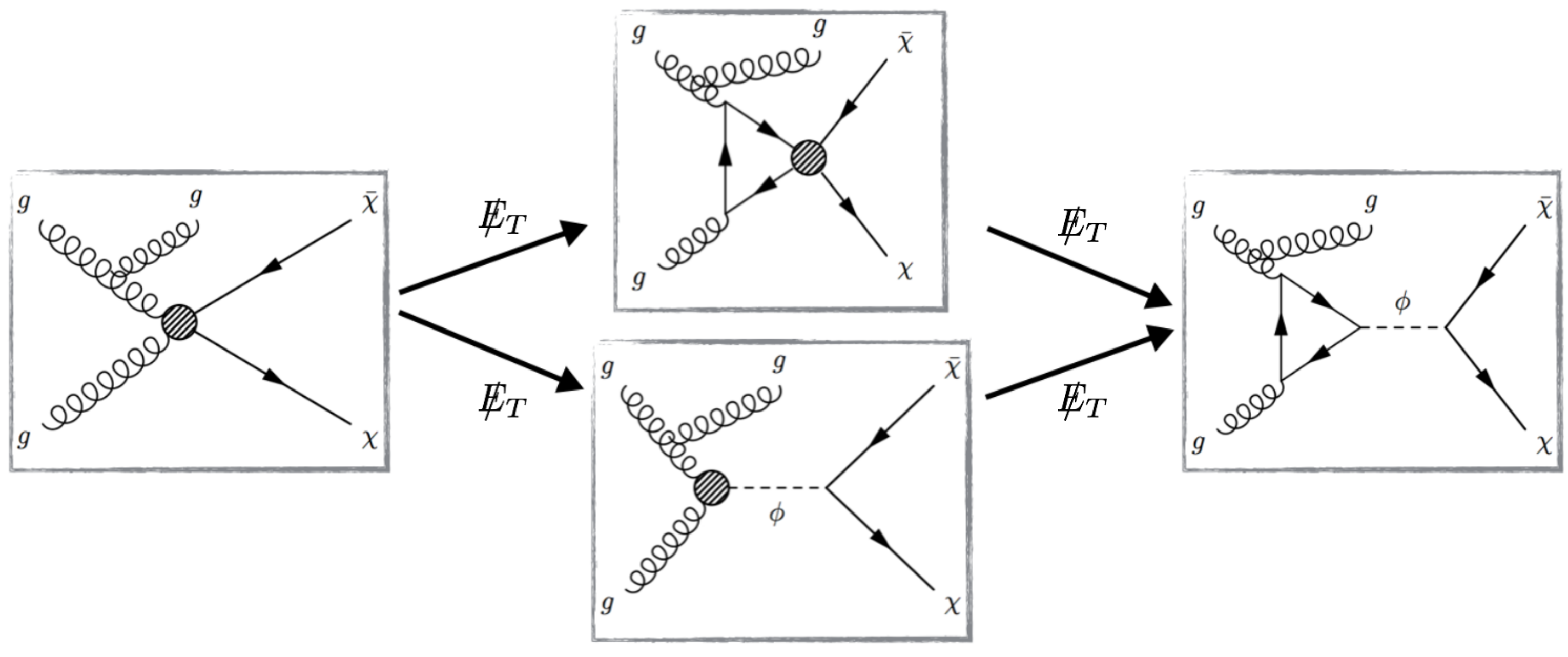}
\caption{A heuristic diagram presenting the successive levels of effective theories that must be expanded as the momentum flow (proportional to the MET) through the interaction increases. 
 On the left we have the EFT $\mathcal{O}_G=\alpha_s/\Lambda^{3}\,\bar{\chi}\chi G_{\mu\nu}G^{\mu\nu}$. In the center 
two effective theories with either $(m_\phi \rightarrow \infty, \mbox{finite}~m_t)$ (top) or $(\mbox{finite}~m_\phi,m_t \rightarrow \infty)$ (bottom). On the right the Full Theory with finite $(m_\phi, m_t)$.
\label{fig:momentum_cartoon}}
\end{figure}

As previous works \cite{Haisch:2012kf,Haisch:2013fla,Haisch:2013ata,Busoni:2014sya,Ghorbani:2014qpa,Crivellin:2014qxa}
have pointed out, scalar and pseudoscalar mediator models and EFTs face unique simulation issues at colliders. Making the well-motivated assumption that the mediator couplings to Standard Model fermions proportional to the Higgs Yukawas, the mediator is primarily produced at the LHC through a loop-induced interaction with gluons. As was noted in the context of scalar EFTs, this loop-induced coupling must be calculated assuming large momentum transfer, as the trigger requirements at the LHC for most dark matter searches require significant transverse momentum in the event. Just as large momenta requires the expansion of a point-like dark matter-Standard Model EFT interaction to include a mediator, the mediator-gluon interaction must also be resolved as the momentum transfer increases $p_{T\phi}=\mathcal{O}( 2m_t)$. A sketch of the successive levels of effective theories is shown in Figure~\ref{fig:momentum_cartoon}. As we will show, the large momentum transfer at the LHC forces us to fully resolve the top-loop induced coupling, just as it forces us to resolve the mediator in the EFT. 

In this paper, we provide two benchmark models for scalar and pseudoscalar mediated simplified models, with a five-dimensional parameter space.  We demonstrate the non-negligible effects of resolving the mediator loop-induced coupling to gluons in collider simulations, compared to the effective interactions. We derive bounds on these parameters using data from direct and indirect detection, as well as predictions assuming that the dark matter is a thermal relic. We then show the existing constraints on these benchmarks from a number of Run-I LHC searches, including -- but not limited to -- the MET plus jets searches that have been of primary interest previously. This comprehensive set of bounds on scalar mediators has not been previously collected, and underlines the necessity of multiple complimentary channels when searching for dark matter at the LHC \cite{Lin:2013sca}. 

In Section~\ref{sec:models} we set up our two benchmark models for scalar and pseudoscalar mediators. We introduce a set of parameters which describe the relevant phenomenology for current and future experimental results. In this section we also show the effects of the resolved top-loop on the distribution of transverse momentum at colliders. In Section~\ref{sec:noncollider} we show constraints on these models from non-collider physics: direct and indirect detection, as well relic abundance cross section. Constraints from existing LHC Run-I missing energy searches are discussed in Section~\ref{sec:collider_bounds} in three channels: missing transverse energy with associated jets, with associated top quark pairs, and  with associated bottom quarks. We apply our constraints to the special case of the 125~GeV Higgs as the scalar mediator in Section~\ref{sec:higgs}.
We then conclude by outlining additional searches and improvements that could be made for future analyses.

\section{Simplified Models} \label{sec:models}

In this paper we consider interactions between Dirac fermion dark matter $\chi$ and 
Standard Model fermions mediated by either a new scalar $\phi$ or a new pseudoscalar $A$. 
Our choice of fermionic dark matter is somewhat arbitrary; our results would translate to
the scalar dark matter case with minor modifications, though this assumption would introduce
additional parameters. Our two benchmark models take the form

\begin{eqnarray}
{\cal L}_S & = & {\cal L}_\text{SM}+ \frac{1}{2} (\partial_\mu \phi)^2 - \frac{1}{2} m_\phi^2 \phi^2 
	+i \bar{\chi} \slashed{\partial} \chi - m_\chi \bar{\chi} \chi  - g_\chi \phi \bar{\chi}\chi 
	- \sum_{\text{fermions}} g_v \frac{y_f}{\sqrt{2}} \phi \bar{f}f  \;, \label{eq:Lphi} \\
{\cal L}_A & = & {\cal L}_\text{SM}+\frac{1}{2} (\partial_\mu A)^2 - \frac{1}{2} m_A^2 A^2 
	+i \bar{\chi} \slashed{\partial} \chi - m_\chi \bar{\chi} \chi - i g_\chi A \bar{\chi}\gamma^5 \chi 
	- \sum_{\text{fermions}} i g_v \frac{y_f}{\sqrt{2}} A \bar{f} \gamma^5f . \label{eq:LA}
\end{eqnarray}
Here, ${\cal L}_\text{SM}$ is the Lagrangian of the Standard Model. Such models introduce five free parameters: dark matter mass $m_\chi$, mediator mass $m_{\phi}$ or $m_A$, the dark matter-mediator coupling $g_\chi$, the flavor-universal Standard Model-mediator coupling $g_v$, and
the mediator width $\Gamma_{\phi}$ or $\Gamma_A$.\footnote{If referring to both the scalar and pseudoscalar models simultaneously, we will use mediator mass $m_{\phi(A)}$ and mediator width $\Gamma_{\phi(A)}$.} Keeping the width as a free parameter leaves open the possibility that the mediator has other 
couplings to additional particles, perhaps in an expanded dark sector. Furthermore, as the cross section for dark matter production, annihilation, and scattering to nucleons is proportional to product of the couplings  $\left( g_\chi g_v \right)^2$ and the width depends on the sum of terms proportional to $g_\chi^2$ and $g_v^2$ separately, 
 by keeping the width as a free parameter, we can set
limits on the combination $g_\chi g_v$ as a function of the width without specifying the individual couplings $g_v$ and $g_\chi$. This is how we will present our bounds in Sections~\ref{sec:noncollider} and \ref{sec:collider_bounds}. 

We set the fermion couplings proportional to the SM 
Yukawa couplings, using the Minimal Flavor Violating (MFV)  assumption~\cite{D'Ambrosio:2002ex}. This avoids introducing precision 
constraints from flavor measurements. Additionally, note that the left-handed Standard Model fermions are $SU(2)_L$ doublets and the right-handed fermions are singlets, while the dark matter  cannot be primarily an $SU(2)_L$ multiplet with $Y\neq 0$, due to direct detection bounds. If $\chi$ is a complete Standard Model gauge singlet, then the mediator $\phi$ or $A$ must have some mixing with the Higgs sector to interact with both the doublet fermions and the dark matter, justifying the Yukawa-proportional coupling assumption. Another possibility is that dark matter is a doublet-singlet mixture, as in the case of a neutralino, allowing the mediator to be an $SU(2)_L$ doublet while still avoiding direct detection constraints. This again involves mass terms in the dark sector proportional to the electroweak symmetry breaking scale, which suggests (though does not require) couplings proportional to Yukawa terms.

We assume that the coupling $g_v$ is universal across all the families of quarks and leptons. One could loosen this requirement without introducing large flavor violation. Taking a cue from two-Higgs doublet models for example, the up-type and down-type couplings could be varied independently. We will not explore this possibility in detail here, but we note such deviations from the baseline model would change the ratios of expected signals in the various collider channels we consider. This again motivates a broad set of experimental searches.

As we have seen, this set of simplified models has some obvious connections with the Higgs sector \cite{Fox:2011pm,Djouadi:2012zc}. As 
a gauge-singlet scalar, the mediator $\phi$ will generically mix with the neutral Higgs. If the SM Higgs is part 
of an extended Higgs sector, then the pseudoscalar $A$ would fit easily into the model (for example, as the
pseudoscalar in a two-Higgs doublet model). If the models are so intimately related to Higgs physics, one might
expect some coupling to $W$ and $Z$ bosons, which we do not allow in our baseline models. We justify this omission by noting that even for scenarios where the scalar and/or pseudoscalar are part of a Higgs sector, deviations from alignment in 
supersymmetry are constrained to be small \cite{Craig:2013hca,Carena:2013ooa}, which in turn implies that the coupling to $W/Z$ bosons of new scalars and pseudoscalars in the Higgs sector 
would likely be small compared to the $125$~GeV Higgs.

Similarly, we would expect explicit dimension-4 $\phi-h$ or $A-h$ couplings in our Lagrangians Eqs.~\eqref{eq:Lphi} and \eqref{eq:LA}.
In a full UV-complete theory, into which the simplified model presumably fits, these couplings would be set by some unspecified dynamics. In this work, we set them to zero for simplicity, as we did for the $W$ and $Z$ couplings.\medskip

\begin{figure}[t!]
\includegraphics[width=\columnwidth]{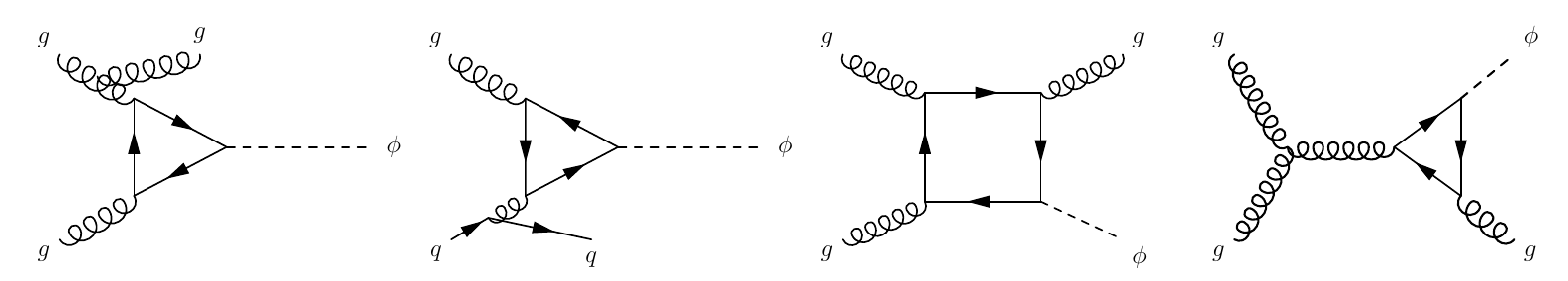}
\caption{Sample of the leading-order Feynman diagrams, in the Full Theory with finite top mass effects, contributing to the scalar plus jet production at the LHC.}
\label{fig:loop_feyndiagram}
\end{figure}

Analogously to the production of the Higgs, the dominant form of dark matter production 
at the LHC would be through gluon fusion, as the tree-level couplings to the light quarks  are Yukawa-suppressed. This production mode is dominantly through the loop induced
$g-g-\phi (A)$ coupling.
Representative diagrams for the leading-jet process are shown in Figure~\ref{fig:loop_feyndiagram}.
Note that in the production of the mediators in channels with associated $b$ or $t$ quarks is largely 
dominated by the tree-level terms, though as in Higgs production, loop effects can be important in the
$\phi(A)+$ heavy flavor channels. 

If the external particles in the loop induced $g-g-\phi (A)$  interaction are on-shell, then it can be exactly calculated 
in a single coupling value, as in Higgs physics.
At leading-order, the on-shell Lagrangians for our two benchmark models gain the additional terms \cite{toploop,pseudoloop1,pseudoloop2,pseudoloop3,Harlander:2005if}
%
\begin{eqnarray}
{\cal L}_{S,\text{loop}} =  \frac{\alpha_s}{8 \pi} \frac{g_v}{v} \tau [1+(1-\tau) f\left(\tau \right)] G^{\mu\nu}{G}_{\mu\nu}\phi
 \;,  \qquad \qquad 
{\cal L}_{A,\text{loop}} =  \frac{\alpha_s}{4 \pi} \frac{g_v}{v} \tau  f\left(\tau \right) G^{\mu\nu}\tilde{G}_{\mu\nu}A
 \;, 
\label{eq:lag_loop}
\end{eqnarray}
%
%
%
where $\tau = 4 m_t^2/m_{\phi (A)}^2$, $y_t$ is the top Yukawa, $v$ is the Higgs vacuum expectation value, and the function $f(\tau)$ is defined as
\begin{eqnarray}
f(\tau)  =  
\begin{cases}
   \arcsin^2 \frac{1}{\sqrt{\tau}} \;,  & \tau \ge 1 \;, \\
   -\frac{1}{4} \left( \log \frac{1+\sqrt{1-\tau}}{1-\sqrt{1-\tau}} - i \pi \right)^2\;, & \tau < 1\;.
\end{cases}  
\end{eqnarray}
We should emphasize that the effective coupling
approximation can be accurately calculated for arbitrary top and mediator masses. However,  for associated production of $\phi$ or $A$
plus jets at collider, with momenta and energy scales
where the loop induced top contributions start to be resolved, that is $p_{T,\phi} = \mathcal{O}(2m_t)$, 
this effective operator breaks down and the one-loop dynamics should be taken into account. Also note that the scalar coupling to gluons is suppressed relative to the pseudoscalar by $\gtrsim30\%$ for mediator masses below $\sim 400$~GeV. This will result in slightly weaker bounds on the scalar model relative to pseudoscalars in channels where the gluon coupling dominates ({\it i.e.}, LHC monojets).

\begin{figure}[t!]
\includegraphics[width=0.4\columnwidth]{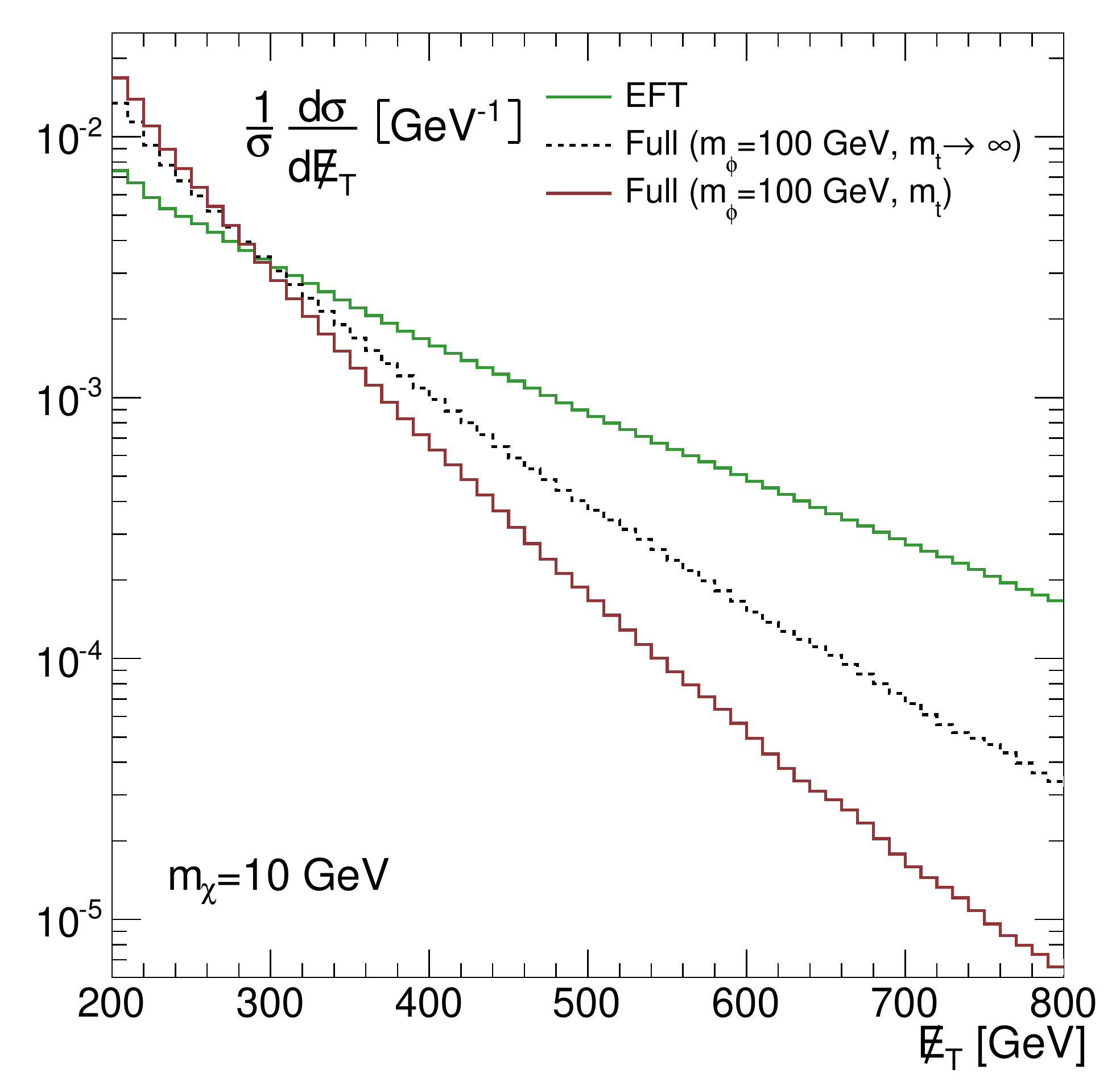}
~~~~~~~~~
\includegraphics[width=0.4\columnwidth]{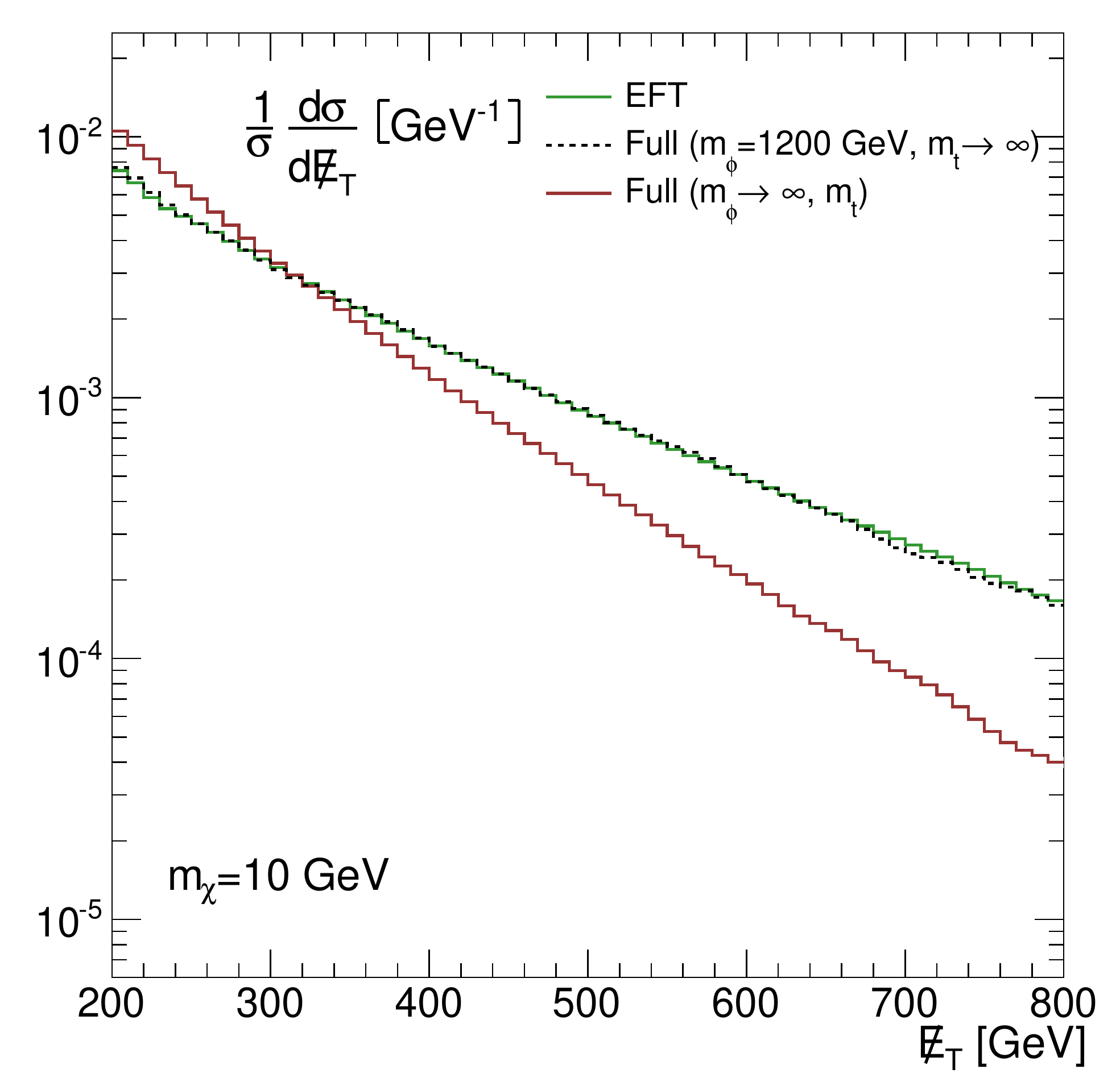}
\caption{Missing energy distribution for the process $pp\rightarrow \bar{\chi}\chi + j$ in the EFT
$\mathcal{O}_G=\alpha_s/\Lambda^{3}\,\bar{\chi}\chi G_{\mu\nu}G^{\mu\nu}$ (equivalent to the left panel of Fig.~\ref{fig:momentum_cartoon}), 
for a finite mediator mass with an effective coupling to gluons $m_t \rightarrow \infty$ (lower center panel of Fig.~\ref{fig:momentum_cartoon}) and the Full Theory including the top mass effects (right panel of Fig.~\ref{fig:momentum_cartoon}).
On the left panel we display the results for a light mediator and on the right for a very heavy one (equivalent to the upper center panel of Figure~\ref{fig:momentum_cartoon}). These distributions were generated at the parton level with {\tt MCFM} and LHC at 8~TeV.} 
\label{fig:etmiss_intro}
\end{figure}

In Section~\ref{sec:collider_bounds} we will discuss further details of the missing transverse energy searches with associated jets used the LHC 
experiments. For this section, it is sufficient to state that significant transverse missing 
momentum is required (that is, large transverse momentum of the $\phi$ or $A$), along with large momentum of at least one
jet, in order to pass the trigger and selection criteria. In events without additional heavy flavor tagging, the primary 
production vertex for the $\phi$ or $A$ will be through the top-loop coupling to gluons, in association with a hard emission
of initial state radiation, see Figure~\ref{fig:loop_feyndiagram}.

In Figure~\ref{fig:etmiss_intro}, we show the missing transverse momentum distribution (MET or $\slashed{E}_T$) for $p p \to \bar{\chi} \chi+ j$ at the 8~TeV LHC, setting $m_\chi = 10$~GeV. Following our sketch (in Figure~\ref{fig:momentum_cartoon}) of the inclusion of integrated-out particles as we resolve effective operators, we present the differential MET distribution from dark matter production for three
different interaction hypothesis: 
\begin{enumerate}
\item for the direct production through an EFT interaction with gluons, $\alpha_s/\Lambda^{3}\,\left[\bar{\chi}\chi G_{\mu\nu}G^{\mu\nu}\right]$; 
\item for the production via a scalar mediator with an effective $g-g-\phi$ interaction vertex, as in Eq.~\eqref{eq:lag_loop}. For comparison purposes, we show both a light (100~GeV) on-shell mediator and very heavy (1200~GeV) mediator which gives dark matter through off-shell production; and
\item  for the production via a scalar mediator where the top-loop has been taken into account via the exact one-loop computation. We show once more a very light  ($m_\phi=100$~GeV) and a very heavy ($m_\phi \rightarrow \infty$) mediator scenarios.  
\end{enumerate}
All these distributions were generated using {\tt MCFMv6.8} \cite{Campbell:2010ff,hj}, where we have extended the process implementation ${pp\rightarrow H(A)+j \rightarrow \tau^+ \tau^-+j}$ in {\tt MCFM}  to accommodate the off-shell mediator production and decay to a dark matter pair. The hard scales are defined as $\mu_F^2=\mu_R^2=m_{\phi(A)}^2+p_{Tj}^2$. For further details on the event generation see Section~\ref{sec:collider_bounds}.

From Figure~\ref{fig:etmiss_intro}, we observe that for heavy mediators above $\mathcal{O}(1~\mbox{TeV})$ and $m_t \rightarrow \infty$ the {\it Simplified Model} can be well described by the EFT. However, for light mediators ($m_\phi=100$~GeV) or finite top mass we see that this approximation breaks down. Moreover, if accurate conclusions about such models are to be drawn from LHC data, it is clearly necessary to include the mediator-gluon interaction (induced by the heavy-quark loops) when the characteristic energies are above $\mathcal{O}(2m_t)$. 

At every stage of returning the integrated particles to the spectrum (as pictorially presented in Figure~\ref{fig:momentum_cartoon}), we see significant changes in the differential cross sections. There is a large decrease in the tail of the MET distributions as first the mediator and then the top-loop are correctly taken into account.  Ignoring these effects in the simplified scalar model will lead to an over-prediction of the cross section at the LHC for a given set of parameters, and thus overly strong limits. Furthermore, when using search techniques that rely on detailed knowledge of the kinematic shape ({\it e.g.}~razor variables \cite{Fox:2012ee,Rogan:2010kb,Chatrchyan:2011ek}), it is of course necessary to fully and correctly understand the shape of the signal distributions.\medskip

\begin{figure}[th!]
\includegraphics[width=0.5\columnwidth]{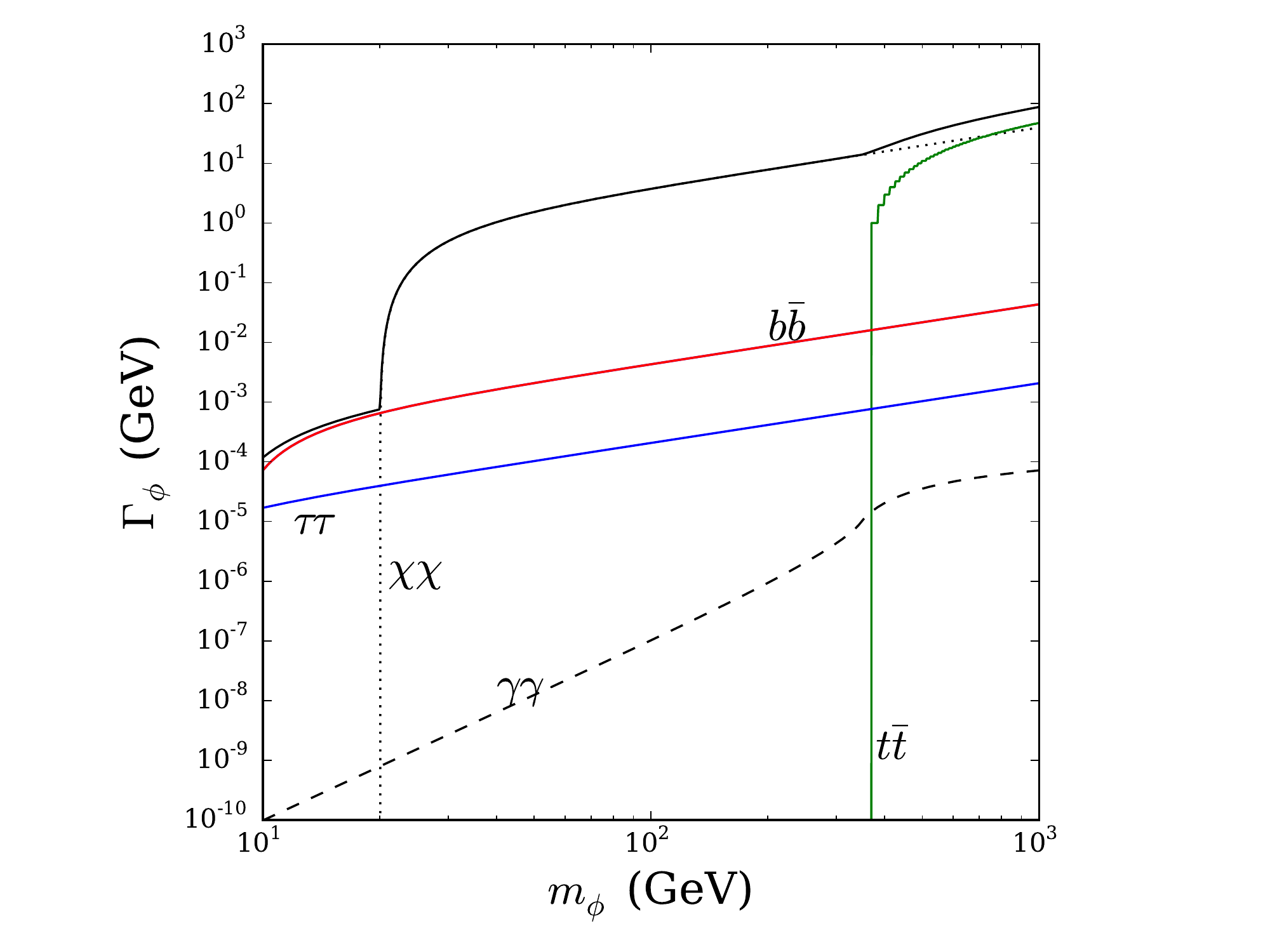}\includegraphics[width=0.5\columnwidth]{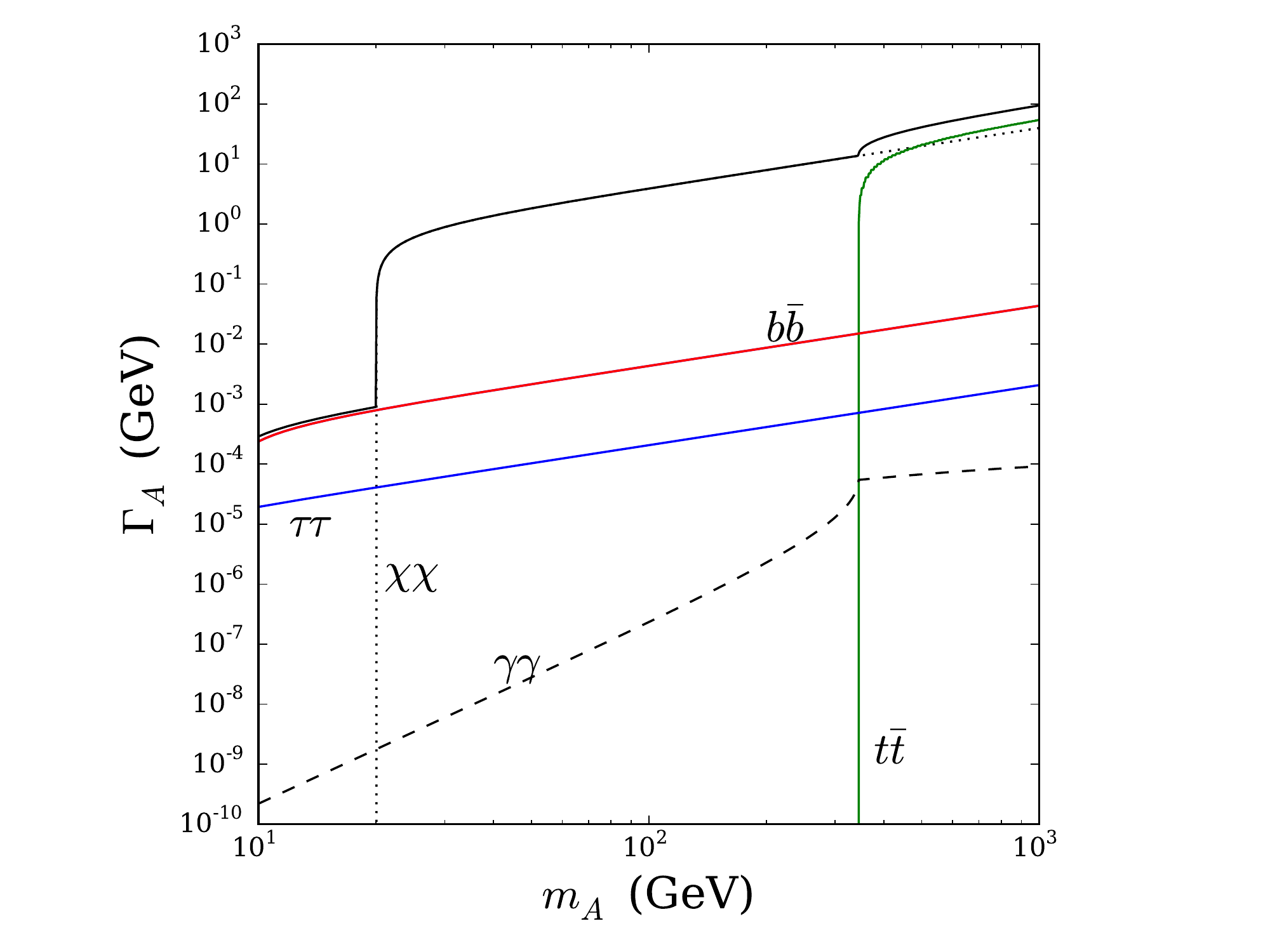}
\caption{The width $\Gamma$ of the scalar $\phi$ (left) and pseudoscalar $A$ (right) decaying into pairs of 10~GeV dark matter (black dotted), top quarks (green), bottom quarks (red), tau leptons (blue), $\gamma\gamma$ (black dashed), and the total width (black solid), as a function of the parent mass $m_\phi$ or $m_A$. Widths are calculated assuming $g_v = g_\chi =1$.   \label{fig:widths}}
\end{figure}

Before moving on to the bounds on the benchmark models, it is useful to consider the widths and branching ratios we might expect in our models of interest. In Figure~\ref{fig:widths}, we show the partial widths for $\phi$ and $A$ decaying into Standard Model particles and dark matter as a function of mass $m_{\phi(A)}$, assuming $m_\chi = 10$~GeV and $g_v = g_\chi = 1$. It is straightforward to rescale the relevant widths if these assumptions are loosened. As can be seen, if $g_v \sim g_\chi$ and $m_\chi \ll m_\phi/2$, the decay of the mediator into dark matter is expected to dominate, unless the mediator is heavy enough for the top channel to open. This is a result of the small Yukawa couplings for the lighter fermions. It is also worth pointing out that differences in rate between the scalar and pseudoscalar  partial decays are given by a distinct scaling pattern with the particle velocity $\beta_\chi=\sqrt{1-4m_\chi^2/m_\phi}$. Namely, the scalar presents a  stronger suppression $\Gamma_{\phi\rightarrow\chi\chi} \propto \beta_\chi^3$ when compared to the pseudoscalar,  $\Gamma_{A\rightarrow\chi\chi} \propto \beta_\chi$. As a result, when the dark matter mass is close to the kinematic limit $2m_\chi \sim m_{\phi(A)}$, we should expect constraints on the couplings of scalars to be weaker than those placed on the couplings to pseudoscalars. When the dark matter is much lighter than the mediator, the coupling constraints on the two models should be equivalent, as in this regime $\beta^{3} \sim \beta \sim 1$. 

\section{Non-Collider bounds} \label{sec:noncollider}

In this section, we derive bounds on our benchmark model parameters, using direct and indirect detection experimental results, as well as the thermal relic abundance calculation. These bounds are complimentary to those set by colliders, which we will consider in Section~\ref{sec:collider_bounds}. However, we caution that care must be taken in extrapolating bounds between different classes of experiments, as there are both particle physics and astrophysical assumptions that must be kept in mind. For example, the direct detection limits rely on an assumption about the local dark matter density and velocity distributions, the latter of which is expected to vary from the standard assumptions used in the experimental results \cite{Kuhlen:2009vh,Lisanti:2010qx,Mao:2012hf,Mao:2013nda,Kuhlen:2013tra,Lee:2013wza,Bozorgnia:2013pua}. While it is possible to some degree to disentangle the astrophysical uncertainties to place limits on the fundamental parameters \cite{Fox:2010bu,Fox:2010bz,Fairbairn:2012zs,Pato:2012fw,DelNobile:2013cta,Feldstein:2014gza}, we cannot lose sight of the assumptions that went into the analysis. Similarly, the parameters that are required to obtain a thermal relic abundance can be changed significantly if additional particles (beyond the minimal set in our benchmark simplified models) are present in the spectrum, or if the flavor-universal assumption for the coupling $g_v$ is lifted. Furthermore, we have no direct knowledge that the dark matter is a thermal relic.

Thus, we wish to emphasize that no single result presented here should be taken as the final word on the limits for our models, since these searches -- along with those of the colliders -- are complimentary and approach the problem from different angles. Despite the caveats, these limits are useful in that they provide a sense of the size of the parameters which might be necessary to obtain a viable model of dark matter, and allow us to focus on regions where particular classes of experiments may dominate.

\subsection{Direct Detection}
Direct detection experiments measure the recoil energy from WIMP-nucleus scattering, placing an upper limit on the dark matter-nucleon elastic scattering cross section. This, like all the bounds we discuss in this paper, requires coupling the dark and visible sectors, and so limits on the scattering cross section provide a constraint on the combination of couplings $g_\chi g_v$. The pseudoscalar model has no velocity or momentum independent scattering cross section with protons and neutrons,  and so has no significant limits from direct detection. However, assuming Dirac dark matter, the scalar mediator induces a spin-independent cross section and so the model parameters are constrained by a number of experiments. The strongest bounds at present come from LUX \cite{Akerib:2013tjd} for $m_\chi \gtrsim 6$~GeV and, at lower dark matter masses, by CDMS-lite \cite{Agnese:2013jaa}.

The fundamental Lagrangian parameters are translated into dark matter-nucleon scattering cross sections using
\begin{eqnarray}
\sigma_{\chi-p,n} & = & \frac{\mu^2}{\pi} f_{p,n}^2, \\
f_{p,n} & = & \sum_{q=u,d,s} f_q^{p,n}\frac{m_{p,n}}{m_q} \left(\frac{g_\chi g_v y_q}{\sqrt{2}m_\phi^2} \right) + \frac{2}{27} f_\text{TG}^{p,n} \sum_{q=c,b,t} \frac{m_{p,n}}{m_q} \left(\frac{g_\chi g_v y_q}{\sqrt{2}m_\phi^2} \right),
\end{eqnarray}
where $\mu$ is the dark matter-nucleon reduced mass, and the parameters $f^{p,n}_q$ and $f^{p,n}_\text{TG}$ are proportional to the quark expectation operators in the nucleon. These must be extracted from lattice QCD simulations \cite{Belanger:2008sj,Young:2009zb,Toussaint:2009pz,Giedt:2009mr,Fitzpatrick:2010em}, and we adopt the values from Ref.~\cite{Fitzpatrick:2010em}. For the purposes of this paper, there is no significant difference between the proton and neutron $f_{p,n}$, and so our dark matter scattering is essentially isospin-conserving. 

\begin{figure}[h]
\includegraphics[width=0.6\columnwidth]{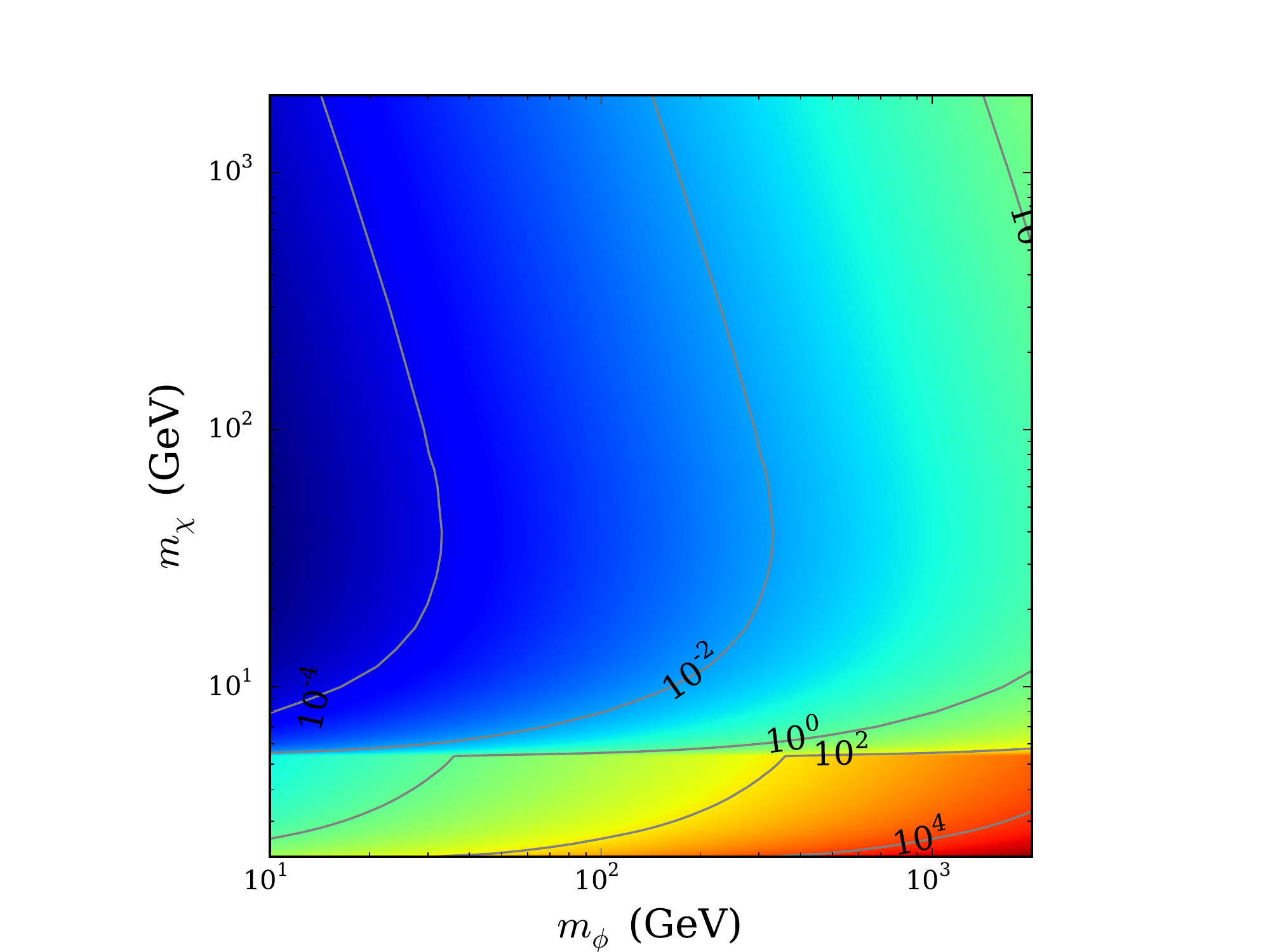}
\caption{Contour plot of 95\% CL upper bounds on the coupling combination $g_\chi g_v$ from LUX \cite{Akerib:2013tjd} and CDMS-lite~\cite{Agnese:2013jaa} direct detection searches on the scalar mediator benchmark model as a function of the mediator mass $m_\phi$ and dark matter mass $m_\chi$. \label{fig:direct_bounds}}
\end{figure}

The finite width is not relevant to these constraints (barring widths of order $m_\phi$), so the bound is placed on the combination $g_\chi g_v$ as a function of dark matter and mediator masses, independent of width. In Figure~\ref{fig:direct_bounds}, we show the upper limits placed by LUX and CDMS-lite at the 95\% confidence level (CL) on the coupling combination $g_\chi g_v$, as a function of the scalar mediator and dark matter masses. The discontinuity visible at $m_\chi \sim 6$~GeV is a result of the sharply weakening LUX bounds being overtaken by the CDMS-lite constraint. As we will continue to do throughout this paper, we include limits on the combination of couplings well above the perturbativity bound $g_\chi g_v \gtrsim4\pi$. Clearly, such enormous couplings are not part of a sensible perturbative quantum field theory. We include them for completeness, and to allow some comparison of the sensitivity of the different classes of experiments. 

%
\subsection{Indirect Detection}

Indirect detection searches look for dark matter annihilating to Standard Model particles in the Universe today. Such processes could be seen by finding an otherwise unexplained excess of gamma rays or positrons coming from an area of expected high dark matter density. While direct detection searches place non-trivial limits on scalar mediator models, such models result in thermally averaged cross sections $\langle \sigma v\rangle$ which are proportional to $v^2$.  The velocity $v$ of dark matter today is very small $\lesssim 10^{-2}c$, and so scalar mediators do not result in significant signals in indirect searches. The velocity-averaged annihilation cross section into Standard Model fermion final states for our two benchmark models are \cite{Buckley:2013jwa}
\begin{eqnarray}
\langle\sigma v\rangle(\chi\bar{\chi} \to \phi^* \to f\bar{f}) & = & \sum_{f} N_f \frac{3g_\chi^2 g_v^2 y_f^2 (m_\chi^2-m_f^2)^{3/2}}{8\pi m_\chi^2\left[ (m_\phi^2 - 4m_\chi^2)^2 + m_\phi^2 \Gamma_\phi^2 \right]} T \\
\langle\sigma v\rangle(\chi\bar{\chi} \to A^* \to f\bar{f}) & = & \sum_{f} N_f \frac{g_\chi^2 g_v^2 y_f^2}{4\pi \left[ (m_A^2 - 4m_\chi^2)^2 + m_A^2 \Gamma_A^2 \right]} \left [m_\chi^2 \sqrt{1-\frac{m_f^2}{m_\chi^2}} +\frac{3m_f^2}{4 m_\chi \sqrt{1-\frac{m_f^2}{m_\chi^2}} } T \right]
\end{eqnarray}
Here, $N_f$ is the number of colors of the fermion $f$, and $T$ is the temperature of the dark matter. As $T\propto v^2$, of our two simplified models, only the pseudoscalars have a thermal annihilation cross section with a velocity-independent term. Thus,  only the pseudoscalar mediator gives significant annihilation in the Universe today with non-trivial bounds set by indirect detection.

Of particular interest, due to their sensitivity to multiple decay channels, are indirect searches for gamma-ray annihilation, either from direct annihilation (resulting in gamma rays with a characteristic energy of $E_\gamma = m_\chi$), or from a cascade of Standard Model decays after annihilation into heavy, charged, and unstable Standard Model particles, which provide a continuum of gamma rays. For gamma-ray energies (and thus dark matter masses) below approximately a TeV, the Fermi Gamma-Ray Space Telescope (FGST) provides the best bounds at present \cite{Abdo:2010ex,GeringerSameth:2011iw,Ackermann:2011wa,Geringer-Sameth:2014qqa}. In particular in this paper we will use the bounds set by the FGST in Ref.~\cite{Ackermann:2011wa}, searching for dark matter annihilation in dwarf spheroidal galaxies orbiting the Milky Way (see also Ref.~\cite{GeringerSameth:2011iw} for an independent analysis). At the moment these are  the most constraining.

We comment that there is an excess of gamma rays from the Galactic Center reported in the FGST data-set \cite{Goodenough:2009gk,Hooper:2010mq,Hooper:2011ti,Boyarsky:2010dr,Abazajian:2012pn,Hooper:2012sr,Hooper:2013rwa,Gordon:2013vta,Huang:2013pda,Abazajian:2014fta,Daylan:2014rsa}. Though the source of these gamma rays is still uncertain \cite{Abazajian:2010zy,Wharton:2011dv,Hooper:2013nhl,Carlson:2014cwa}, if interpreted in terms of dark matter, it could be be accommodated by annihilation through a pseudoscalar mediators with Standard Model couplings proportional to Yukawas~\cite{Abdullah:2014lla,Basak:2014sza,Berlin:2014pya,Arina:2014yna,Cheung:2014lqa,Balazs:2014jla}, as in our benchmark simplified model.

\begin{figure}[b!]
\includegraphics[width=0.59\columnwidth]{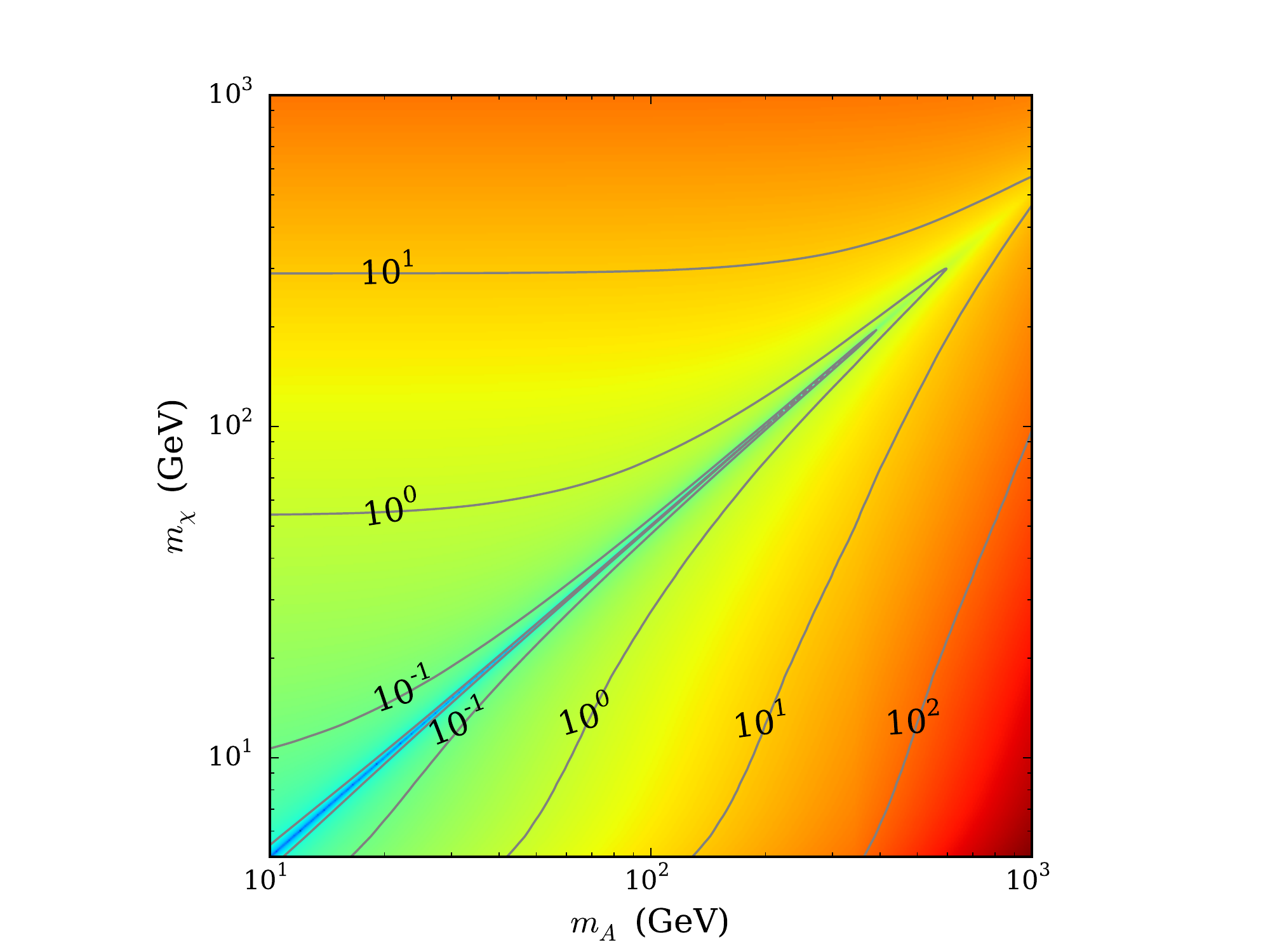}
\caption{Contour plot of 95\% CL upper bounds on $g_\chi g_v$ derived from indirect detection constraints set by the FGST dwarf spheroidal analysis \cite{Ackermann:2011wa} in the $b\bar{b}$ channel,  as a function of the pseudoscalar mediator mass $m_A$ and the dark matter mass $m_\chi$. The width is set assuming $g_v = g_\chi$, which is relevant only near resonance. \label{fig:indirect_bounds}}
\end{figure}

In this paper, we use only the 95\% CL upper limits on the indirect annihilation cross section into pairs of $b$-quarks from the FGST dwarf analysis \cite{Ackermann:2011wa}, converted to limits on our model parameters by calculating the velocity averaged cross section $\langle \sigma v\rangle$ (see Ref.~\cite{Buckley:2013jwa} for details) evaluated at $v \to 0$. Constraints on $g_\chi g_v$ are shown in Figure~\ref{fig:indirect_bounds}. The width $\Gamma_A$ can play an important role here near resonance, so to reduce the parameter space we choose a width under the assumption that the two couplings are equal. This has only a minor effect on the majority of the parameter space. We further assume that no other annihilation channels are present.

\subsection{Thermal Relic Abundance}

By measuring CMB anisotropies, surveys such as the Planck mission have measured the dark matter contribution to the Universe's energy budget to be $\Omega_{\chi}h^2 = 0.1187 \pm 0.0017$ \cite{Ade:2013zuv}. From standard Boltzmann relic density calculations \cite{Kolb}, this implies a thermal annihilation cross section of $\langle \sigma v\rangle \sim 3 \times 10^{-26}$~cm$^3$/s. 
If we assume that $\phi$ is the only connection between the dark and visible sectors, and we further assume that the dark matter is a thermal relic, we can calculate the couplings $g_\chi$ and $g_v$ necessary for the production of the observed density of dark matter. 

As with indirect detection, near resonance ($m_\phi \sim 2 m_\chi$) we must assume knowledge of the mediator width $\Gamma_{\phi(A)}$. We make the same assumption as before: that the width is calculated as if $g_v = g_\chi$. Annihilation near resonance can have significant effects on the cross section during thermal freeze-out, which we take into account using the methods outlined in Ref.~\cite{3Excpns}. Away from resonance, the thermally averaged cross section becomes identical to that calculated for the indirect detection constraints, evaluated at the freeze-out temperature $T_f = m_\chi/x_f \sim m_\chi/ 25$.

Additionally, when $m_\phi < m_\chi$, dark matter can annihilate in the process $\bar{\chi}\chi \to \phi \phi$, followed by decay of the $\phi$. Thus a thermal relic can be obtained even when $g_v \sim 0$, as long as the $\phi$ is not sufficiently long-lived as to decay after Big Bang Nucleosynthesis. Such detector-stable particles are completely consistent as a dark matter mediator, but may require searches targeted towards displaced vertices. For the purposes of this paper, will not consider these models in more detail here, though the possibility should not be ignored. 

\begin{figure}[h!]
\includegraphics[width=0.42\columnwidth]{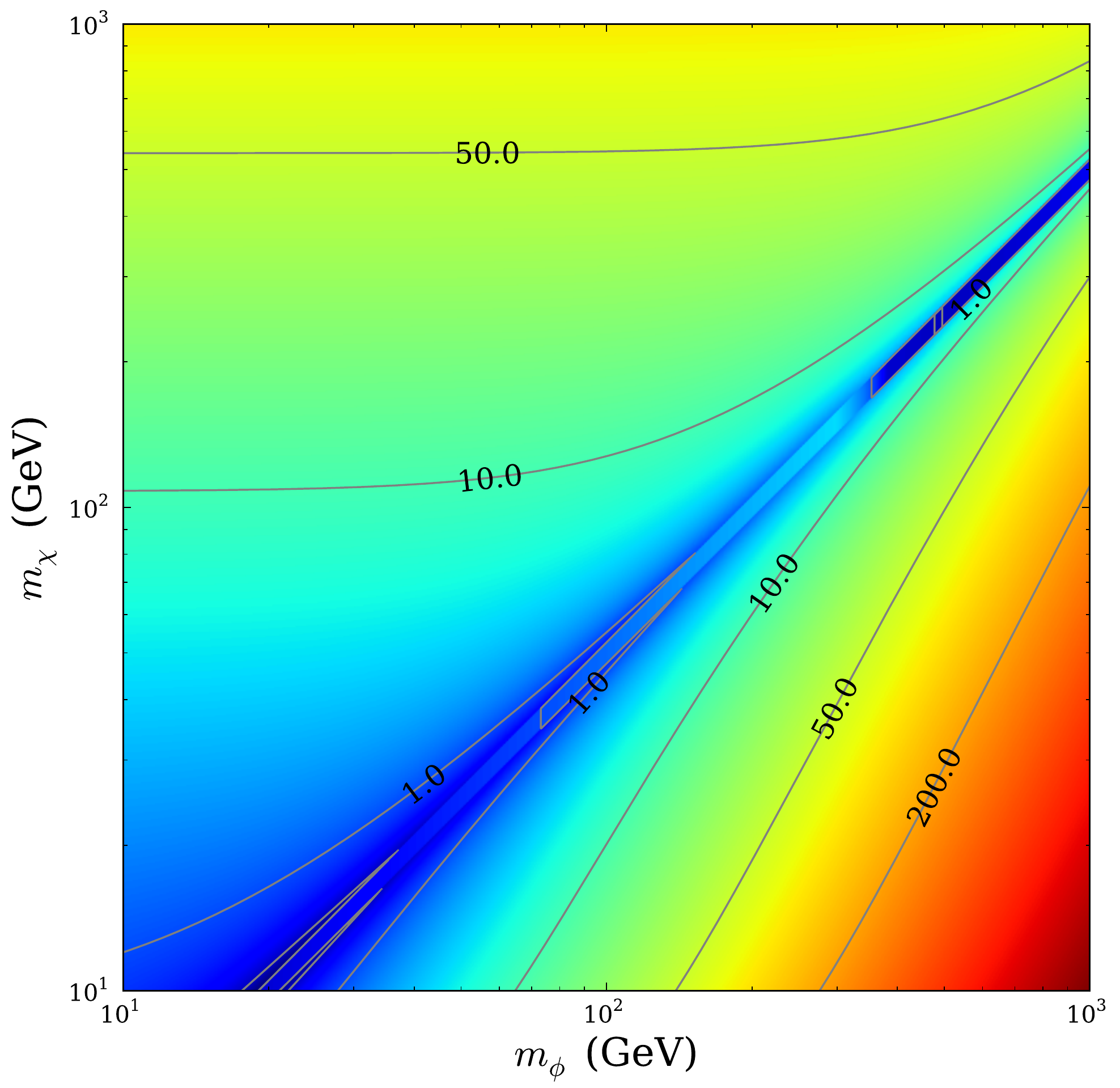}
~~~~~
\includegraphics[width=0.42\columnwidth]{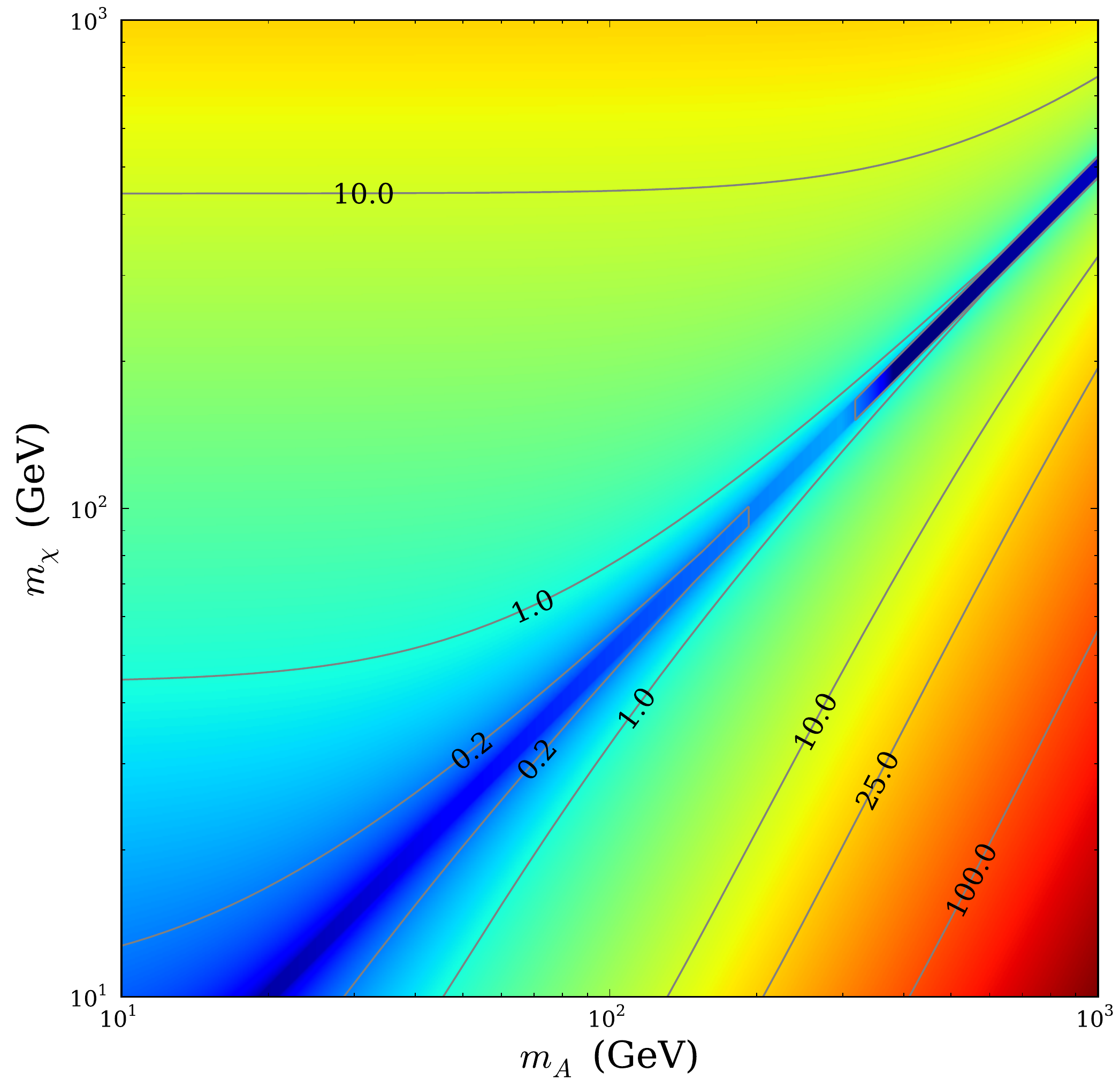}
\caption{Required values of $g_\chi g_v$ as a function of mediator mass $m_{\phi(A)}$ and dark matter mass $m_\chi$ assuming that dark matter is a thermal relic and the only annihilation channel is $\bar{\chi}\chi \to \phi(A) \to \bar{f}f$, for the scalar (left) and pseudoscalar (right) simplified models. \label{fig:thermal_bounds}}
\end{figure}

The required combinations of couplings $g_\chi g_v$ in order to obtain a thermal abundance are shown in Figure~\ref{fig:thermal_bounds}, assuming the only open channel is $\bar{\chi}\chi \to \phi(A) \to \bar{f}f$. We again emphasize that the regions of mass and coupling parameter space that do not yield a correct thermal relic under our specific set of assumptions are still of great interest, and so these constraints should not be taken as the final word on dark matter physics. Recall that we are discussing a simplified scenario, which presumably fits into a larger model of the dark sector. If the couplings are too small to give the correct relic abundance, then the simplified model predicts an over-abundance of dark matter from thermal processes. However, entropy dilution could reduce the dark matter density, if the physics in the Early Universe is non-standard \cite{Hooper:2013nia}. Somewhat more prosaically, the full theory of the dark sector could contain additional mediating particles that increase the annihilation cross section \cite{ArkaniHamed:2008qn}. If the couplings under consideration are larger than required for thermal annihilation, then non-thermal models of dark matter (such as asymmetric dark matter) are an attractive possibility \cite{Kaplan:2009ag,Cohen:2009fz,Belyaev:2010kp,Davoudiasl:2010am,Buckley:2010ui,Buckley:2011kk}. 

\section{Collider bounds}
\label{sec:collider_bounds}

Having placed bounds on our simplified models from direct detection, indirect detection, and under the assumption that the dark matter obtains the thermal relic abundance, we now turn to bounds from the LHC. The most obvious signature for dark matter at colliders is missing transverse momentum (more colloquially, missing transverse energy). When dark matter is produced it escapes the detector unseen, leaving an imbalance of momentum which can be measured in the transverse plane. This missing transverse momentum is a powerful signature for new physics models. MET signatures must be accompanied by some associated production of visible particles, both for momentum conservation and triggering. We consider three signatures in this paper: MET with associated untagged jets, MET with two associated dileptonic tops, and MET plus one or two $b$-tagged jets. 

In all these searches, we follow our previous policy of setting upper bounds on the combination $g_v g_\chi $. However, unlike the previous examples, the branching ratios of the mediators $\phi$ or $A$ are integral to the bounds set. By setting the limit on the combination of couplings, the mediator width $\Gamma_{\phi(A)}$, which depends on $g_\chi^2$ and $g_v^2$ separately, must be specified as an independent parameter.

Both the simplified models and EFTs can consider scenarios where the mass hierarchy is inverted ($2m_\chi > m_{\phi(A)}$). For EFTs, this makes no difference (other than bringing into question the applicability of the effective operator approach). However, in our simplified models, if the mediator is light enough to be produced at a collider, but the dark matter is heavy enough so that it cannot be the product of on-shell decay of the mediator, then it is likely that better search strategies would be those based around the decays of the mediator into visible final states. For heavy mediators ({\it i.e.} $m_{\phi(A)} \gtrsim 1$~TeV at the LHC) the searches for dark matter with masses satisfying $2m_\chi < m_{\phi(A)}$ would be reliant on the off-shell mediator production. For scalars and pseudoscalar mediators, however, the current constraints in this regime from the LHC turn out to be extremely weak. As a result, in this paper, we will concentrate on the $m_{\phi(A)} > 2m_\chi$ regime, and leave the remainder of the mass plane for future work. \medskip
 
Considering the importance of the width on the collider constraints for much of the accessible parameter space, we chose to parametrize the derived limits on $g_\chi g_v$ at fixed dark matter and mediator masses, varying the width $\Gamma_{\phi(A)}$. We choose two mediator masses: ${m_{\phi(A)} = 100}$~GeV, and 375~GeV, and $m_\chi = 40$~GeV. For on-shell mediator production, the bounds could be easily extrapolated to other dark matter masses (up to the kinematic limit $2m_\chi = m_{\phi(A)}$) by rescaling the overall branching ratio into dark matter at a new mass point. Recall that the kinematic suppression for scalars ($\beta^3$) will be more significant than that of pseudoscalars ($\beta$) for the 100~GeV benchmark, as a 40~GeV dark matter particle is near the kinematic threshold.

\subsection{Mono-jet Search}
At a hadron collider, unless the mediator has large couplings to $W/Z/\gamma$ compared its coupling to the colored partons, we would expect the strongest constraints to come from the production of dark matter in association with an initial state jet radiation~\cite{Goodman:2010ku,Beltran:2010ww,Fox:2011pm,Goodman:2010yf,Rajaraman:2011wf,Fox:2012ee}. Both ATLAS \cite{ATLASmonojet} and CMS \cite{CMSmonojet} have performed dedicated ``monojet'' searches using Run-I LHC data at $\sqrt{s} = 8$ TeV. We note that the ``monojet'' moniker is something of a misnomer, as these analyses do allow a second high-$p_T$ jet in the sample. 

We use results from CMS~\cite{CMSmonojet} to derive bounds on couplings for our benchmark models. The CMS search used a data sample corresponding to an integrated luminosity of $19.5$~fb$^{-1}$. Events are required to have one jet with $p_{Tj} > 110$ GeV. A second jet is allowed, but no more than two jets with $p_{Tj}> 30$~GeV. Signal events are grouped into seven MET bins: $\slashed{E}_T > 250$, 300, 350, 400, 450, 500, and 550~GeV. The CMS Collaboration has provided the number of events in each bin that can be accommodated as signal at the 95\% CL, which we use to place bounds on $g_\chi g_v$ as a function of $m_{\phi(A)}$, $m_\chi$, and $\Gamma_{\phi(A)}$, using the most constraining limit from any of the seven MET signal bins.

As we showed in Section~\ref{sec:models}, the treatment of the $g$--$g$--$\phi(A)$ interaction as an effective operator would introduce significant errors in the extrapolated bounds on the model parameters. Hence, accurate distributions of MET and jet $p_T$ require simulation of $\phi$ or $A$ plus a hard parton including the 
exact heavy-quark loop effects. We implement this in {\tt MCFMv6.8} \cite{Campbell:2010ff,hj}, modifying the process ${pp\rightarrow H(A)+j \rightarrow \tau^+ \tau^-+j}$ in {\tt MCFM} to produce events files which can be subsequently showered and hadronized by {\tt Pythia8}~\cite{Sjostrand:2006za,Sjostrand:2007gs}, then fed into a detector simulator.  Note that, while the CMS analysis allows a second jet, our {\tt MCFM} simulation is limited to one hard parton, though additional jets are generated through the {\tt Pythia8} parton shower. See Refs.~\cite{hjets,Campanario:2010mi, Buschmann:2014twa} for issues pertaining the simulation of the second jet  including the top mass effects. In addition, we generalized the {\tt MCFM} implementation including the possibility of off-shell mediator production. As there are no full Next-to-Leading order (NLO) predictions including the top mass effects  for this process in the literature, we include these effects via a flat correction factor $K\sim 1.6$ obtained using the infinite top mass limit~\cite{kfactor} . Our hard scales are defined as $\mu_F^2=\mu_R^2=m_{\phi(A)}^2+p_{Tj}^2$, and we used the {\tt CTEQ6L1} parton distribution functions \cite{cteq}.
\begin{figure}[t!]
\includegraphics[width=0.4\columnwidth]{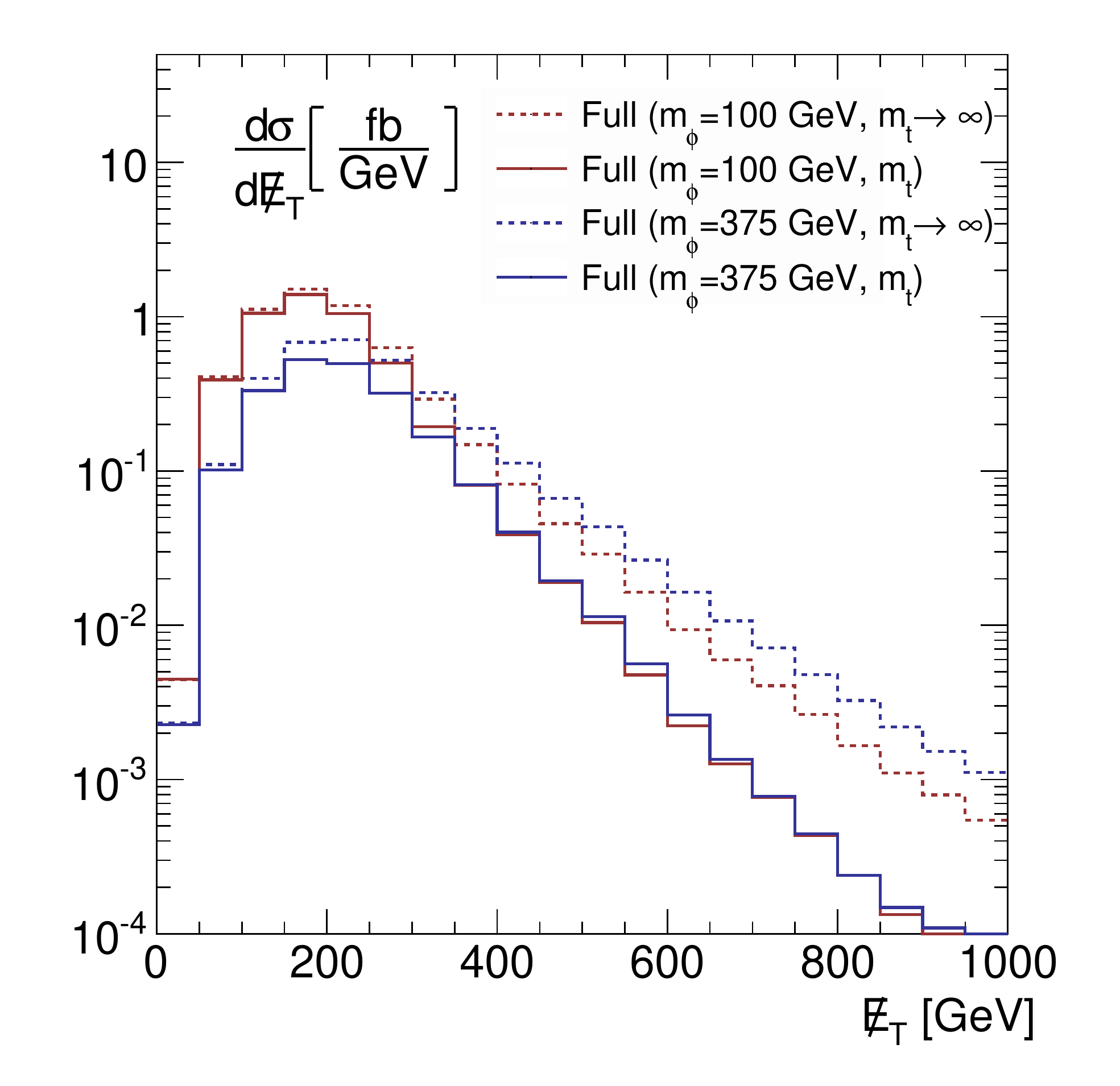}
\includegraphics[width=0.4\columnwidth]{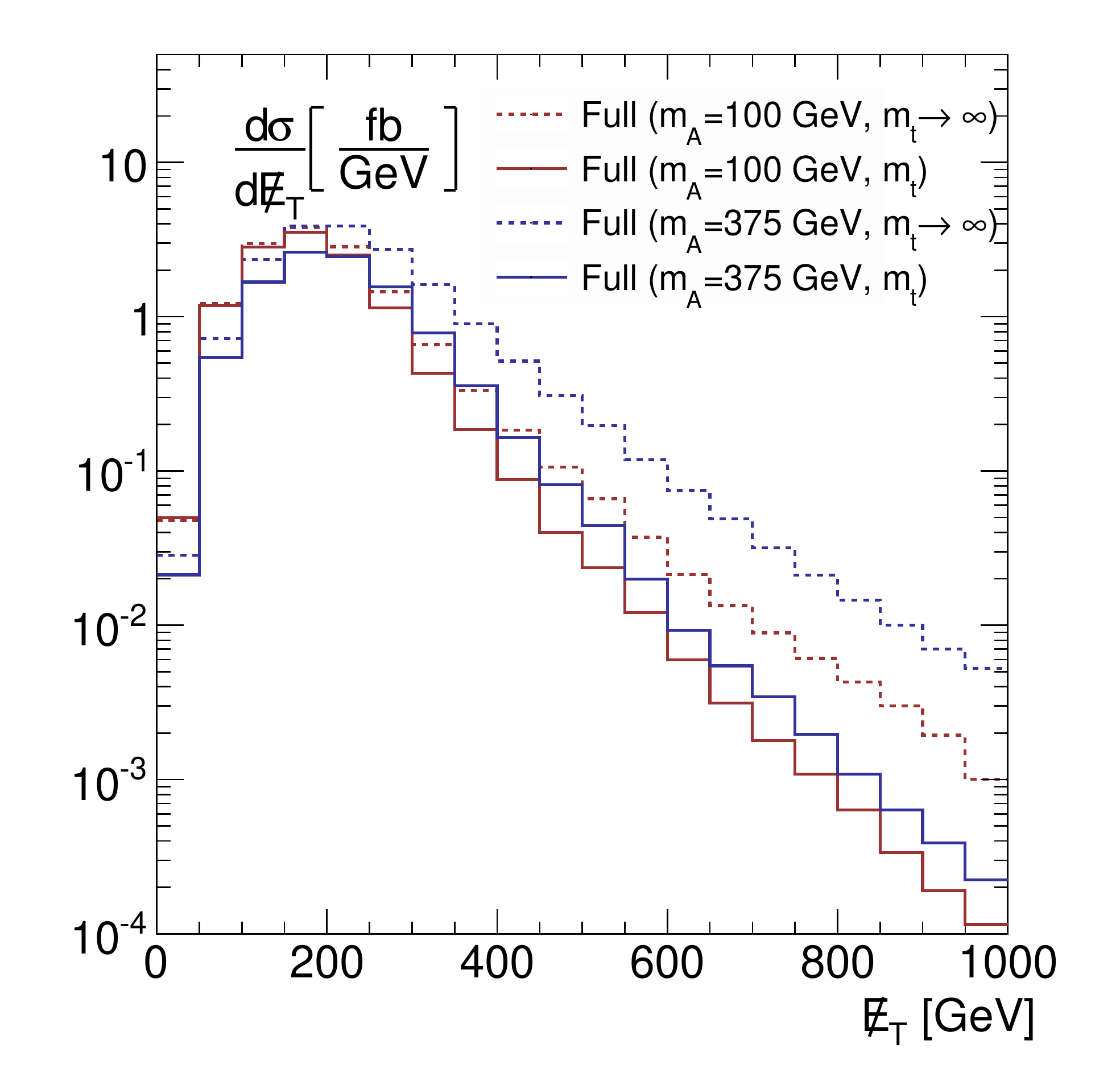}
\caption{Missing transverse momentum  differential cross sections for the scalar (left panel) and pseudoscalar (right panel) mediators.  The leading order effective gluon couplings are shown as dashed lines, and the exact loop-induced calculations are solid. We assume the LHC at 8~TeV.
\label{fig:pt_distributions}}
\end{figure} 

\begin{figure}[b!]
\includegraphics[width=0.48\columnwidth]{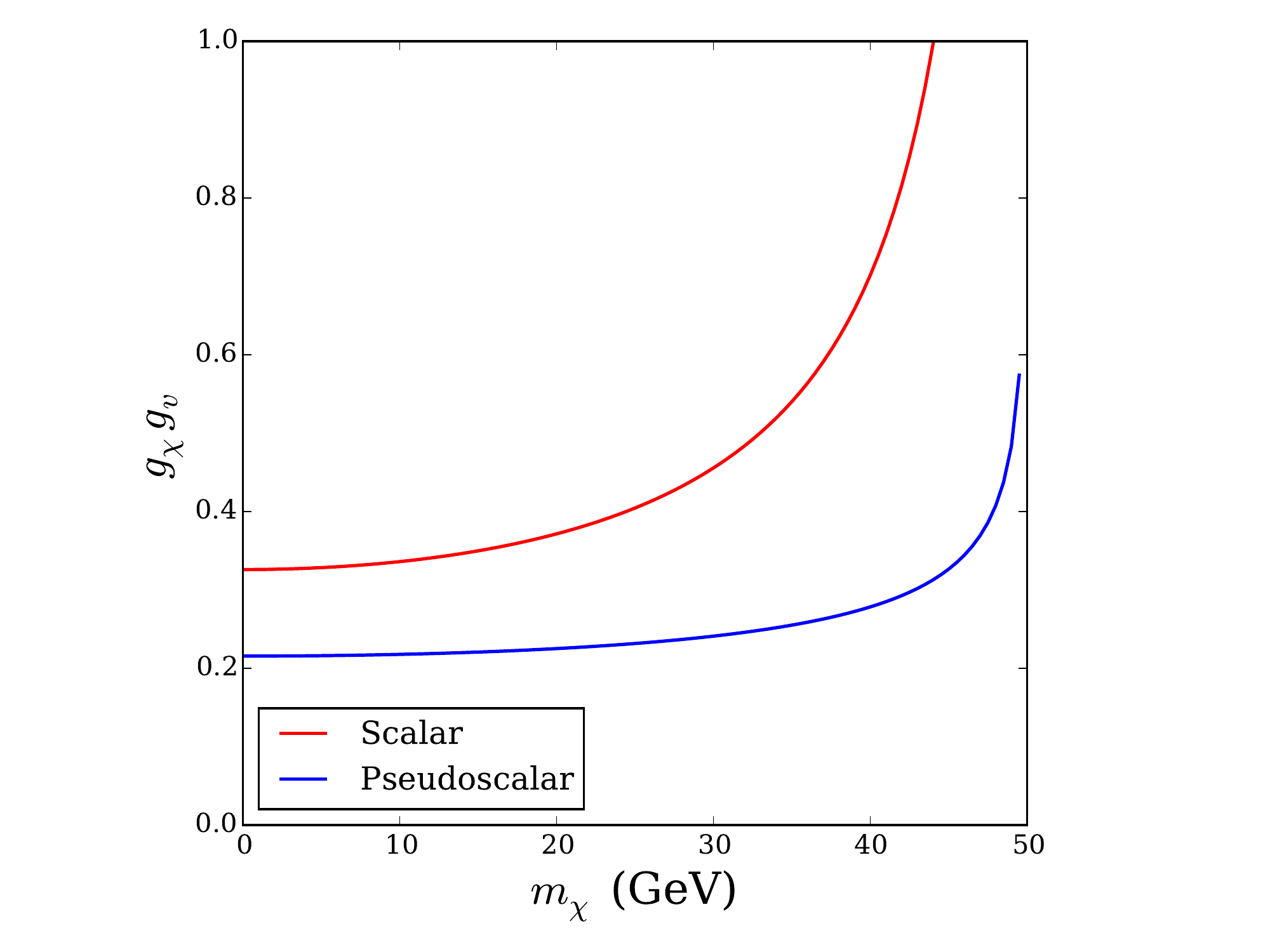}
\caption{Lower limit on the coupling $g_\chi g_v$ set by the CMS monojet search as a function of dark matter mass $m_\chi$, assuming mediators of 100~GeV, $\Gamma_{\phi(A)}/m_{\phi(A)} = 10^{-3}$, and exclusively on-shell production of dark matter. The constraint on the scalar mediator is shown in red and pseudoscalars in blue.  \label{fig:dm_dependence}}
\end{figure}

Whereas the primary effect on the bounds placed on the combination $g_\chi g_v$ from varying the width $\Gamma_{\phi(A)}$ is just a rescaling of the branching ratio to dark matter, there can be small secondary effects when the width is significant compared to the mediator mass. To investigate these effects, as well as demonstrate the importance of the full simulation on the bounds, we also generate dark matter events in our two simplified models using {\tt MadGraph5}~\cite{Alwall:2011uj,Alwall:2014hca}.
This implementation starts with the inclusion of our {\it Simplified Model}, presented in Eq.~\eqref{eq:lag_loop}, into {\tt Feynrules}~\cite{Alloul:2013bka} which generates a model file that is subsequently used by {\tt MadGraph}.  In {\tt MadGraph} we produce $\phi(A)$ events matched up to two jets  
via the MLM scheme~\cite{Mangano:2006rw}. We also include the detector simulation through {\tt Delphes3}~\cite{deFavereau:2013fsa}. In Figure~\ref{fig:pt_distributions}, we compare the distributions for the leading jet $p_T$ and the MET in the narrow width approximation generated by both {\tt MCFM} and {\tt MadGraph5}, after the CMS event selection criteria. As in Figure~\ref{fig:etmiss_intro}, the effective gluon operator overestimates the distribution tails, which would lead to an overly aggressive bound on the couplings. Notice that these differential distributions do not differ from the exact result by just a flat factor, but have different shapes. While these effects are important here, they will be even more critical in future LHC runs, where the energies will be higher and the MET cuts will be harsher. To confirm  the consistency of our implementation, we have produced results in the EFT limit $(m_t\rightarrow \infty,m_\phi\rightarrow \infty)$ and validated it against the CMS EFT bounds~\cite{CMSmonojet}.

Using these simulations, we place 95\% CL bounds on $g_\chi g_v$ as a function of $\Gamma_{\phi(A)}/m_{\phi(A)}$, for 100 and 375 GeV mediators and $m_\chi =40$~GeV. Our results are shown in Figure~\ref{fig:scalar_bounds} for the scalar mediator and Figure~\ref{fig:pseudo_bounds} for the pseudoscalar. Two points from these results should be addressed in detail.
\begin{enumerate}

\item[$i$)] The different dependence on the scalar and pseudoscalar widths on $\beta$ have an important effect on the results. For the light mediator, the scalar partial width into dark matter ($\propto\beta^{3}$) significantly reduces the total cross section when compared to the pseudoscalar ($\propto\beta$). As a result, the couplings to the scalar must be larger than the pseudoscalar for the $100$~GeV mediators. For the heavy mediator, neither scenario has a significant kinematic suppression. This is dependent on our choice of dark matter mass; as the dark matter mass increases, we expect to see the scalar bounds weaker faster than the pseudoscalar. This is explicitly an effect due to on-shell production of the mediator; if the dark matter mass was heavier than $m_{\phi(A)}/2$, then the monojet channel would only be sensitive to production of dark matter via an off-shell mediator, which does not scale with the kinematic suppression factor. In Figure~\ref{fig:dm_dependence}, we show the scaling of the monojet bound as a function of dark matter mass, assuming a 100~GeV scalar or pseudoscalar (and $\Gamma_{\phi(A)}/m_{\phi(A)} = 10^{-3}$). 

\item[$ii$)]  In the case of the {\tt MCFM} results, the changing width only causes a rescaling of the total rate of mediator production times decay into dark matter through the changing branching ratios. While this is the dominate effect for the finite width calculation, there is a subleading effect at $\Gamma_{\phi(A)}/m_{\phi(A)} \gtrsim 0.1$, where the tail of the mediator $p_T$ distribution (and thus the MET) can be increased relative to the narrow width approximation. This is a result of the mediator being able to be produced with $q^2$ very far away from the expected mass, convolved with the proton parton distribution functions. For the 100~GeV mediators, as the width is increased this secondary effect causes the bound on $g_\chi g_v$ to weaken less rapidly than one would expect from the branching ratio alone. The effect is negligible for the 375~GeV mediators.
\end{enumerate}

\subsection{Heavy Flavor Searches}

One would expect that the strongest constraint that the LHC can place on the dark matter decay channels of our benchmark scalar and pseudoscalar mediators comes from the general jets plus missing transverse energy search discussed previously, as the production cross section here is highest. However, channels with missing energy associated with particles other than untagged jets can have significantly lower backgrounds (and different systematics) than the monojets. Therefore, we can and should consider searches in additional channels. Though we will often find that limits placed on the couplings will be weaker than those placed by the monojet search, this approach is still critical as the LHC continues to ramp up to higher energies and luminosities. Recall that we are working with a simplified model, purposefully constructed to minimize the number of free parameters. Therefore, under these assumptions we can predict the exact ratio of signal strength in multiple channels, as the cross section for each is set by the same masses and couplings. However, we must be open to deviations from the simplified model. For example, if the couplings to up- and down-type couplings are not set by a universal coupling $g_v$, or if the loop-induced gluon coupling does not depend solely on the couplings to top and bottom quarks, then it is quite possible that the signal in the monojet channel could be suppressed relative to other production mechanisms. Discovery in more than one channel would also allow better understanding of the theoretical underpinnings of any new physics. \medskip

\begin{figure}[t!]
\includegraphics[width=0.5\columnwidth]{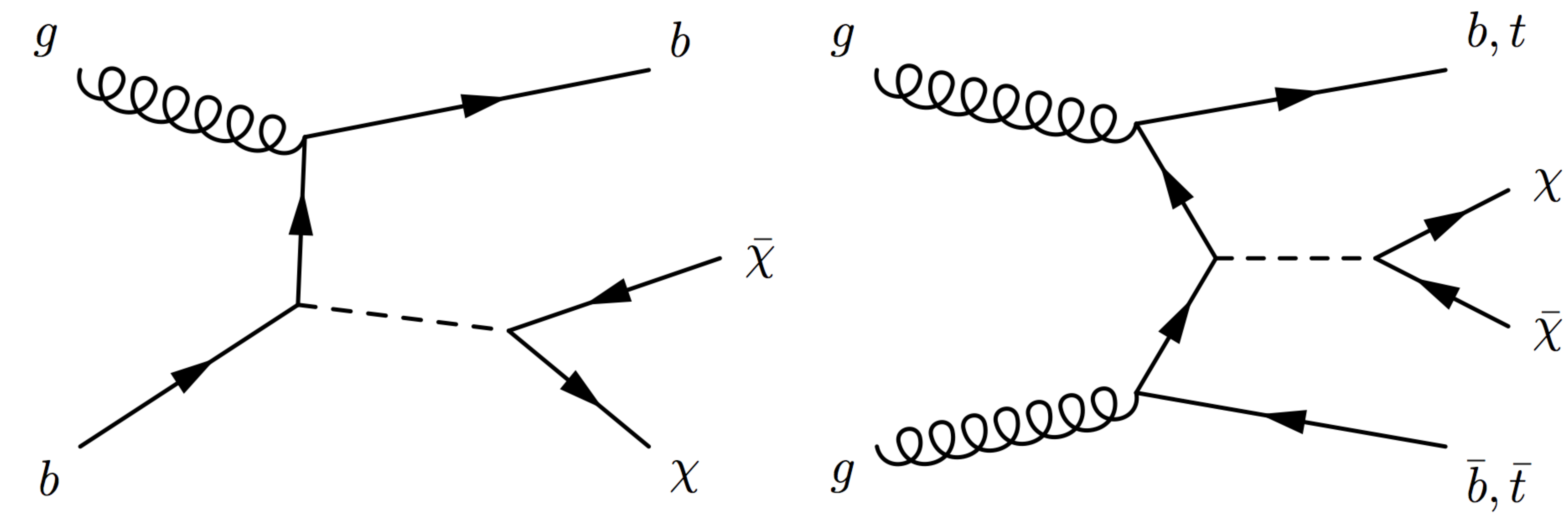}
\caption{Representative Feynman diagrams contributing to heavy quark flavor plus dark matter production at the LHC in our {\it Simplified Models}.  \label{fig:feyn_heavy}}
\end{figure}

With that motivation in mind, it is clearly important to look for new physics in many associated channels. Even when considering modifications to the baseline models, it is still reasonable to assume that the interactions with fermions are largely MFV, and therefore that the mediator is most strongly coupled to the heaviest fermions. Therefore, we show here limits on production of the $\phi$ or $A$ in association with top and bottom quarks, followed by the invisible decay of the mediator into dark matter. Some of the main production diagrams for such processes are shown in Figure~\ref{fig:feyn_heavy}.

\begin{figure}[b!]
\includegraphics[width=0.5\columnwidth]{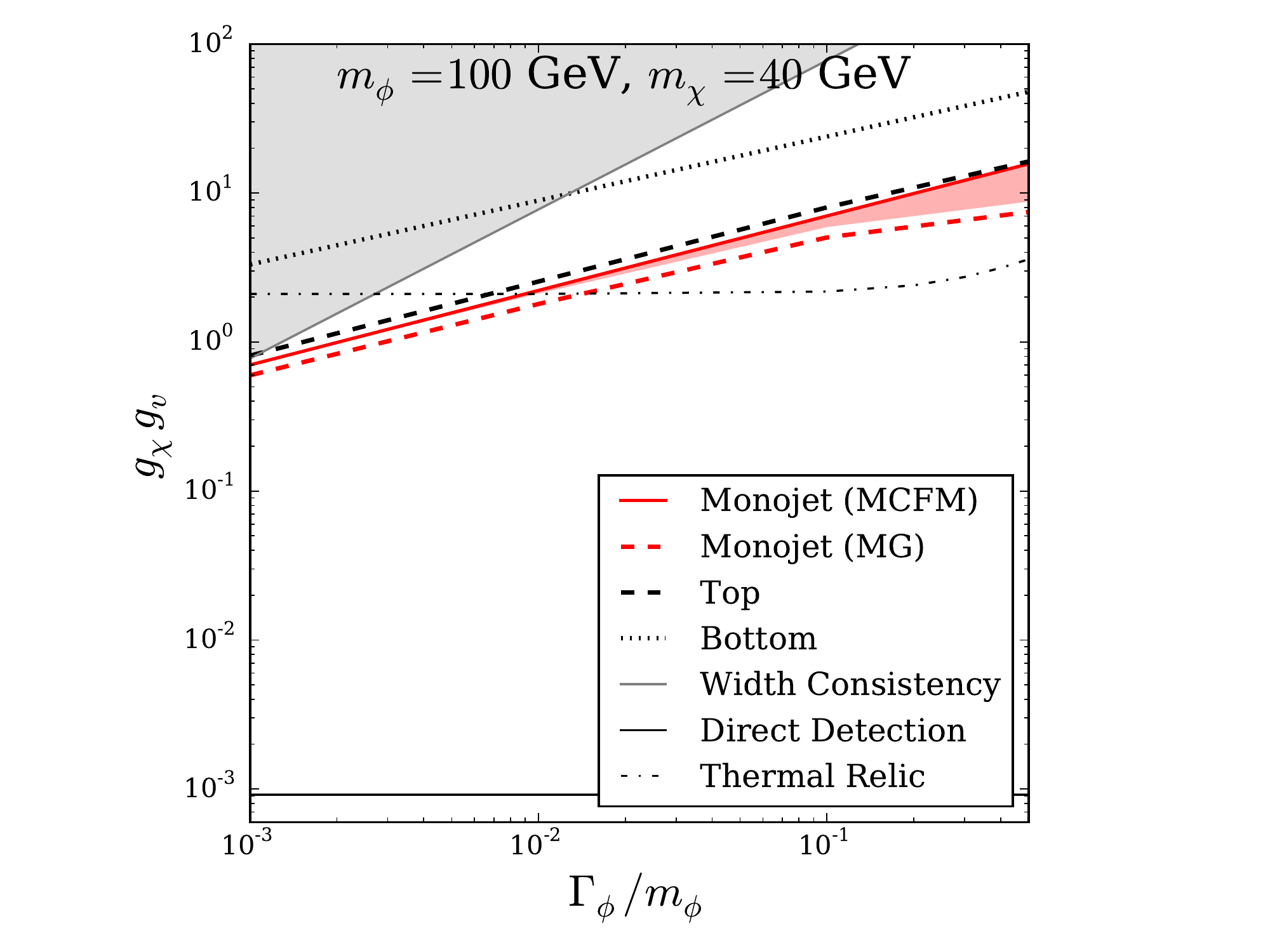}\includegraphics[width=0.5\columnwidth]{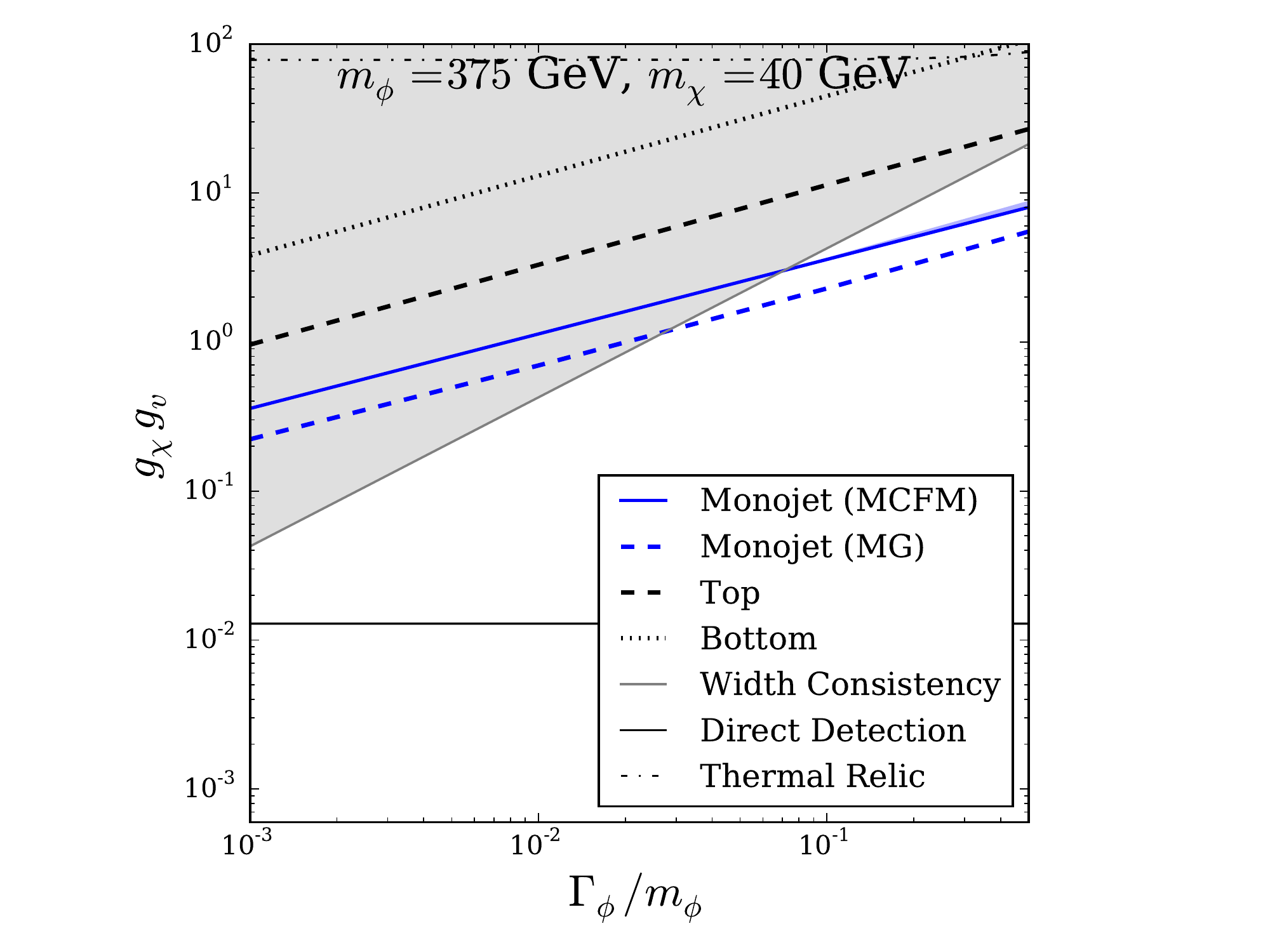}
\caption{95\% CL upper limits on $g_\chi g_v$ for scalar mediators from collider searches as a function of $\Gamma_\phi/m_{\phi}$, assuming 40~GeV dark matter and 100~GeV (left) and 375~GeV (right) scalar mediators. The limit from the CMS monojet 
search is shown as the solid colored (red or blue) line for the Full Theory including heavy quark mass effects {\tt MCFM} calculation. The {\tt MadGraph} effective operator CMS monojet constraint is shown in dashed color. The shaded region indicates an extrapolation of the finite width effects to the {\tt MCFM} results. The constraint from the top pair plus missing energy search is the dashed black line, and the $b$-jet plus missing energy search limit is the dotted black line. The horizontal solid black line shows the direct detection limit from LUX and CDMS-lite. The grayed-out region indicates where the minimum width consistent with $g_\chi g_v$ is greater than the assumed width.
\label{fig:scalar_bounds}}
\end{figure}

\begin{figure}[t!]
\includegraphics[width=0.5\columnwidth]{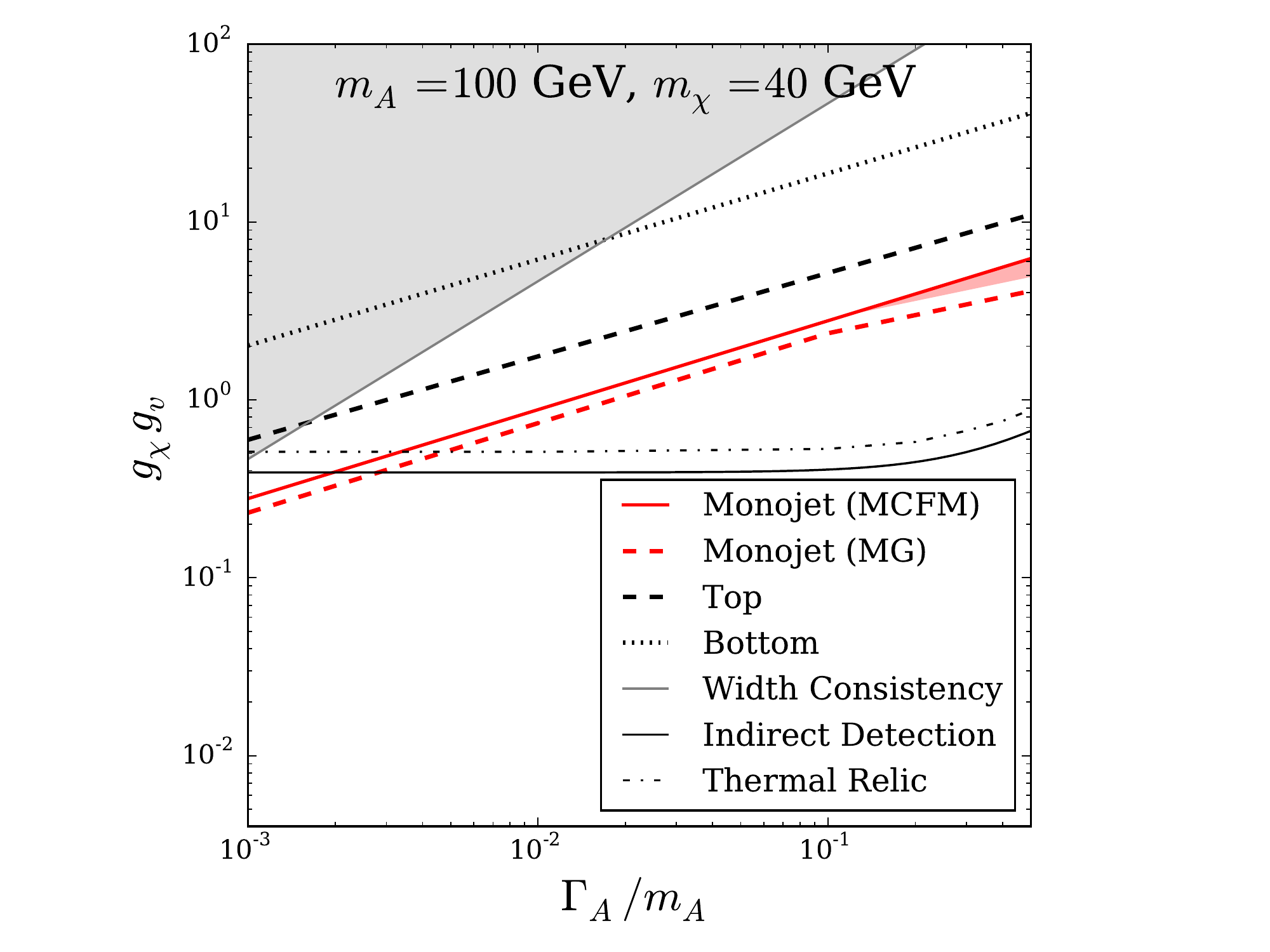}\includegraphics[width=0.5\columnwidth]{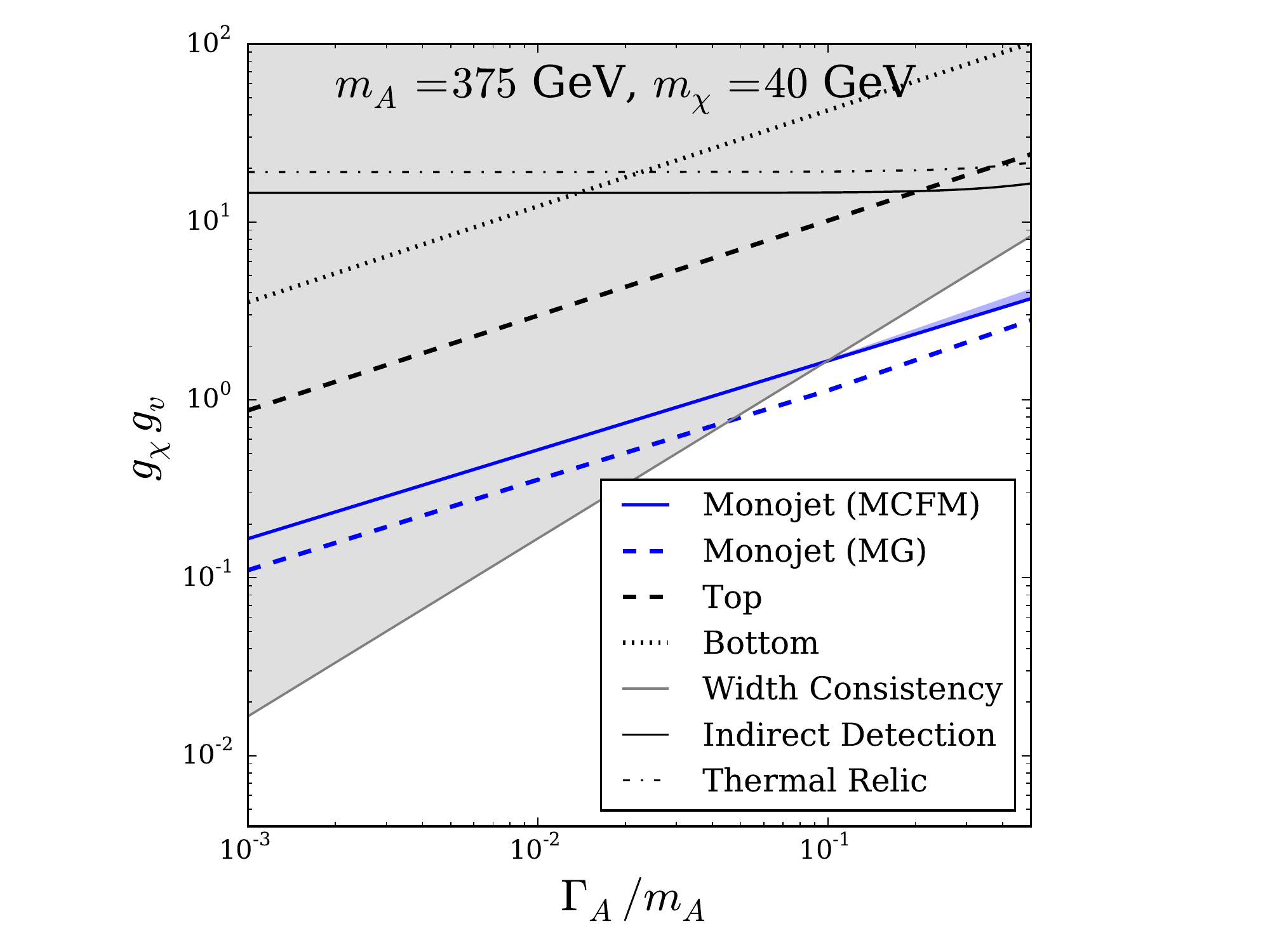}
\caption{95\% CL upper limits on $g_\chi g_v$ for pseudoscalar mediators from collider searches as a function of $\Gamma_A/m_{A}$, assuming 40~GeV dark matter and 100~GeV (left) and 375~GeV (right) pseudoscalar mediators. The limit from the CMS monojet 
search is shown as the solid colored (red or blue) line for the Full Theory including heavy quark mass effects {\tt MCFM} calculation. The {\tt MadGraph} effective operator CMS monojet constraint is shown in dashed color. The shaded region indicates an extrapolation of the finite width effects to the {\tt MCFM} results. The constraint from the top pair plus missing energy search is the dashed black line, and the $b$-jet plus missing energy search limit is the dotted black line. The horizontal solid black line shows the indirect detection limit in the $b\bar{b}$ channel from FGST.
\label{fig:pseudo_bounds}}
\end{figure}

We use the CMS dedicated search for dark matter produced in events with dileptonic tops \cite{CMSttbar}, performed on 19.7~fb$^{-1}$ of integrated luminosity at the 8~TeV LHC.
The analysis requires exactly two isolated leptons with individual $p_T > 20$~GeV and $\sum{p_t} > 120$ GeV, and at least two jets with $p_T > 30$ GeV. The invariant mass of the leptons must be greater than $20$~GeV, and if they are the same flavor, a $Z$-mass veto of $|m_{\ell\ell} - 91~\mbox{GeV}|>15$~GeV is applied. The two jets are required to have invariant mass of less than $400$~GeV. The signal region is $\cancel{E}_T > 320$ GeV. As with the monojet analysis described previously, we can straightforwardly recast the CMS limits to apply to our benchmark models, based on the number of events seen in their signal region. Signal was generated using {\tt MadGraph5}, passed through the {\tt Pythia6} and {\tt Delphes3} pipeline described earlier. As in the monojet case, we validate our results using the dark matter EFT to compare with the CMS results.
We show the bounds from this channel on $g_\chi g_v$ for our benchmark mediator models (for mediators of 100 and 375~GeV, and 40~GeV dark matter) as a function of mediator width in Figures~\ref{fig:scalar_bounds} and \ref{fig:pseudo_bounds}.  \medskip

Finally, we can consider the associated production of the mediator $\phi$ or $A$ with $b$-quarks. Until recently, no dedicated dark matter search similar to the monojet or dileptonic top plus MET analyses has been performed for the process $p p \rightarrow \chi \bar{\chi} +b \bar{b}$, and constraints could only be extracted using the sbottom searches $p p \rightarrow \tilde{b}^{*} \tilde{b} \rightarrow  \chi \bar{\chi} + b \bar{b}$ from CMS \cite{CMSbbar} and ATLAS \cite{Aad:2013ija}.
These searches have selection criteria which are far from ideal for the kinematics of the simplified models, but they do place relevant constraints directly on the tree-level interaction between $b$-quarks and the mediator. 

Recently however, ATLAS has published a dedicated search for dark matter produced in associated with $b$-tagged jets in 20.3~fb$^{-1}$ of 8 TeV data \cite{Aad:2014vea}. Two signal categories in this search are relevant for our analysis here. In both, the analysis vetoes events with leptons that have $p_T > 20$~GeV and requires $\cancel{E}_T > 300$~GeV. The azimuthal angle between all jets and the MET must be $\Delta \phi >1$.  Signal Region SR1 requires one or two jets, at least one of which must be $b$-tagged (at a $60\%$ efficiency) and have $p_T > 100$~GeV. Signal region SR2 requires three or four jets in the event, again requiring at least one to be $b$-tagged with $p_T > 100$~GeV. If a second $b$-tagged jet exists, it must have $p_T> 60$~GeV, and the second highest $p_T$ jet must have $p_T > 100$~GeV. ATLAS provides the 95\% CL upper limit on the number of events in each signal region which can be accommodated by new physics, and we validate our simulation using the EFT results.

We again generate our signal events using {\tt MadGraph5}, through the tree-level coupling of the mediator and the $b$-quarks.  As with the monojet search, for each of our benchmark models, we use the strongest limit on $g_\chi g_v$ set by either of these signal regions.   \medskip

The results from this analysis are shown along with our previous limits as a function of mediator width in Figures~\ref{fig:scalar_bounds} and \ref{fig:pseudo_bounds}. Along with the bounds derived from colliders, we include the direct and indirect constraints (for scalar and pseudoscalar models, respectively) and the required value of $g_\chi g_v$ to obtain the thermal relic abundance. While it is a very useful benchmark to compare the experimental sensitivity, note that coupling values that diverge from that required for a thermal relic are still experimentally and theoretically interesting: as we consider only a {\it Simplified Model}, we do not attempt to specify the full theory. Further, we do not even know that dark matter is in fact a thermal relic. If dark matter was generated through some asymmetric process (like baryons), then one would not expect the low-energy annihilation channels to obtain a thermal abundance.

In Figures~\ref{fig:scalar_bounds} and \ref{fig:pseudo_bounds}, we also show the exclusion region of coupling-width parameter space that is theoretically inconsistent. While we cannot specify a width only from the coupling combination $g_\chi g_v$, we can calculate the minimum possible width (assuming only decays into the dark matter and the Standard Model fermions) that is consistent with a given value of $g_\chi g_v$. That is, for a given width $\Gamma_{\phi(A)}$, we find the minimum value of the product $g_\chi g_v$ which would allow
\begin{equation}
\Gamma_{\phi(A)} > \frac{g_\chi^2 m_{\phi(A)} }{8\pi} \left(1-\frac{4m_\chi^2}{m_{\phi(A)}^2} \right)^{n/2}+\sum_f \frac{g_v^2 y_f^2 m_{\phi(A)}}{16\pi} \left(1-\frac{4m_f^2}{m_{\phi(A)}^2} \right)^{n/2},
\end{equation}
for any values of that $g_\chi$ and $g_v$ which satisfy the product constraint (here $n= 1$ for pseudoscalars and 3 for scalars). We gray-out the regions of $g_\chi g_v$ parameter space where minimum width possible for any $g_\chi$ and $g_v$ is larger than the assumed $\Gamma_{\phi(A)}$. 

Examining Figures~\ref{fig:scalar_bounds} and \ref{fig:pseudo_bounds}, it is interesting to note that the top constraints on the scalar mediator are competitive (within the accuracy of our simulated search) with those of the monojet channel at low mediator masses. This is due to the relative suppression of the scalar coupling to gluons compared to the coupling to the fermions Eq.~\eqref{eq:lag_loop}. The pseudoscalar gluon coupling does not have the same level of suppression, leading to a larger production cross section in the monojet channel, and thus better bounds when compared to the heavy flavor channel. As the mediator mass increases, the production of a heavy particle in association with the two massive tops is suppressed, and the monojet constraint regains its preeminence for the scalar model. 

The $b$-tagged channel places significantly weaker constraints on these models than the monojet or the top channels. However, as this probes directly the coupling to the down-sector, it would be sensitive to deviations the universal coupling assumption in a way that the top channel is not, as the top channel relies on the same coupling as the loop-induced monojet search, unless new colored particles coupling to the mediator exist in the spectrum.

The direct detection constraints are also very powerful compared to the collider reach (though for dark matter masses less than $\sim 6$~GeV, the colliders are more constraining) for scalar mediators, while the pseudoscalars are much less constrained by the indirect searches, are comparable with the current LHC constraints. However, as we argued previously, multiple probes in multiple channels are still necessary, as simple modifications of the basic model or experiment-specific backgrounds and uncertainties could increase the sensitivity of one mode while decreasing another. In our search for new physics, we must exhaust all reasonable search strategies.

\section{Higgs Mediators \label{sec:higgs}}

As we have often mentioned throughout this work, there are obvious connections between our scalar and pseudoscalar simplified models and Higgs physics. In addition to the possible embedding of the simplified models into extended Higgs sectors, the couplings (both tree-level and loop-induced) even in the general scenarios have many similarities with Higgs physics (due in part to the MFV assumption). The correct technique for generation of high $p_T$ events through the gluon-mediator coupling was also inherited from Higgs physics. 

With these considerations, it is reasonable to ask what bounds can be set on the 125 GeV Higgs itself, assuming that it is the scalar mediator between the visible and the dark sector. This is the well-known ``Higgs Portal'' scenario for dark matter \cite{Burgess:2000yq,Davoudiasl:2004be,Patt:2006fw,Andreas:2008xy,Barger:2008jx, Lerner:2009xg,He:2009yd,Barger:2010mc,Djouadi:2011aa,Kanemura:2010sh,Mambrini:2011ik,He:2011de,Han:2013gba,Greljo:2013wja,Okada:2013bna,Chacko:2013lna,deSimone:2014pda,Endo:2014cca,Drozd:2014yla,Englert:2013gz} (similarly, one could consider the ``dilaton'' portal \cite{Bai:2009ms,Agashe:2009ja,Blum:2014jca,Efrati:2014aea}).  

Collider bounds on the 125~GeV Higgs decaying to dark matter can be placed in two ways. First, just as we have done previously, we can place limits on the total cross section from the monojet and heavy flavor channels, which can be translated into limits on the coupling of the Higgs to dark matter. Secondly, we can use the experimental measurements of the Higgs width to constrain the addition of new channels to Higgs decay. 

\begin{figure}[b!]
\includegraphics[width=0.5\columnwidth]{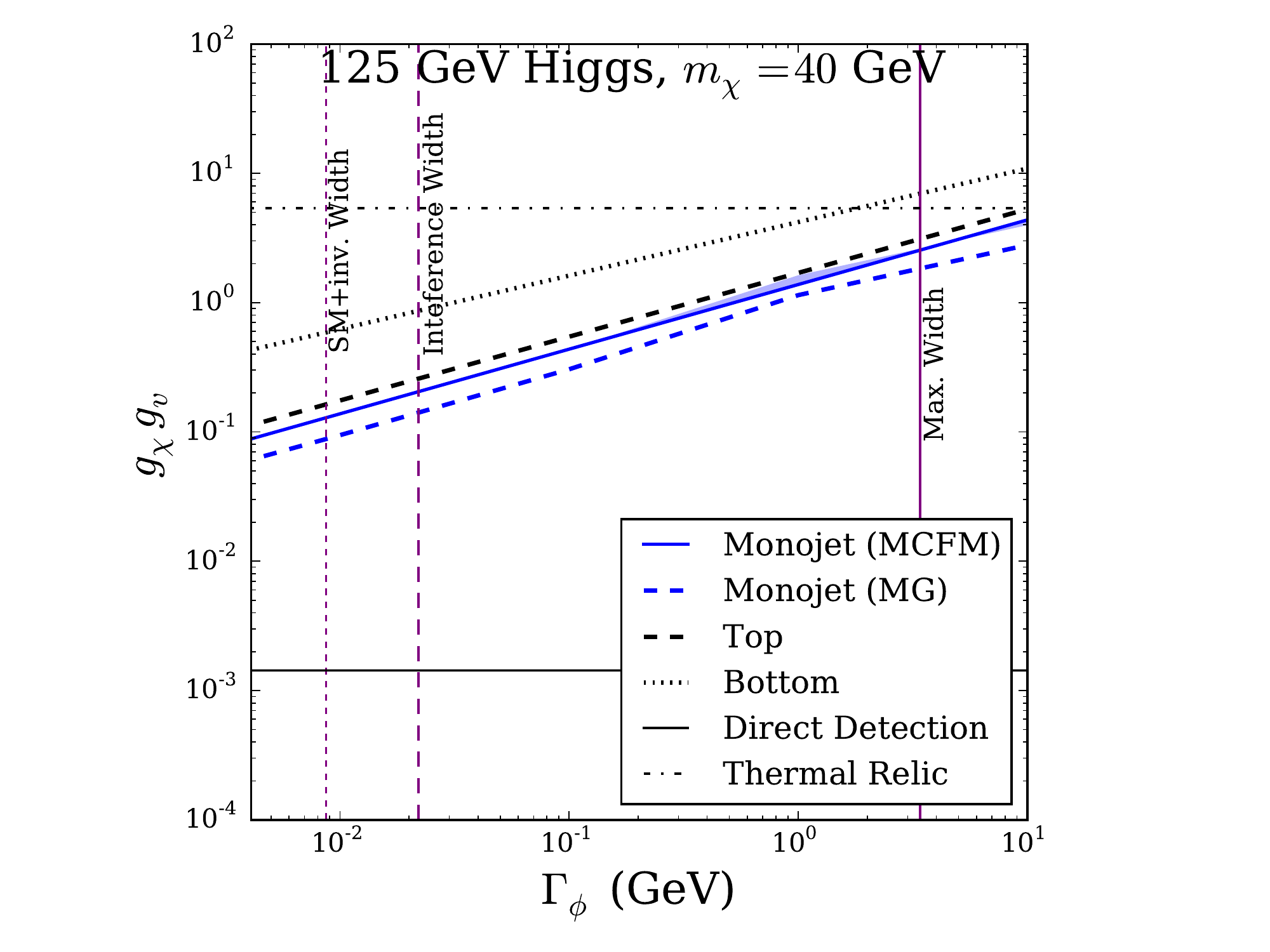}
\caption{95\% CL upper limits on $g_\chi g_v$ for the 125 GeV Higgs from collider searches as a function of the width $\Gamma$, assuming 40~GeV dark matter. The limit from the CMS monojet 
search is shown as the solid colored (red or blue) line for the Full Theory including heavy quark mass effects {\tt MCFM} calculation. The {\tt MadGraph} effective operator CMS monojet constraint is shown in dashed color. The shaded region indicates an extrapolation of the finite width effects to the {\tt MCFM} results. The constraint from the top pair plus missing energy search is the dashed black line, and the $b$-jet plus missing energy search limit is the dotted black line. The horizontal solid black line shows the direct detection limit from LUX and CDMS-lite. Three vertical lines show experimental limits on the 125 GeV Higgs' width assuming Standard Model couplings and an invisible branching ratio of 54\% \cite{Chatrchyan:2014tja}
(dotted purple), the upper limit on the width from interference with the $Z$ \cite{Khachatryan:2014iha}
(dashed purple), and the maximum possible width from the $4\ell$ lineshape (solid purple)  \cite{Chatrchyan:2013mxa}.
\label{fig:higgs_bounds}}
\end{figure}

We can extract constraints on the total width of the Higgs in three different ways. First, if we require that the coupling to the Standard Model is exactly that of {\it the} Standard Model Higgs, then by requiring the visible production and decay channels are consistent with observations, the total invisible branching ratio must be less than $0.54$ at 95\% CL \cite{Chatrchyan:2014tja} (see also Ref.~\cite{Aad:2014iia}). Given the Standard Model Higgs width of 4.1~MeV \cite{Heinemeyer:2013tqa}, the addition of a decay to dark matter saturating this bound gives a total width of at most 8.9~MeV. Furthermore, as this assumes that $g_v = 1$, in this restricted subset of the model space, the dark matter coupling can be constrained to be less than
\begin{equation}
g_\chi^2 \leq \frac{8\pi}{m_h} \left(1-\frac{4m_\chi^2}{m_h^2} \right)^{-3/2} \times \left( 8.9~\mbox{MeV} \times 0.54 \right). 
\end{equation}

This chain of logic does require that the Higgs couplings be exactly the Standard Model values. Somewhat weaker constraints can be placed 
on the invisible branching ratio once this assumption has been lifted. This does not extend to statements about the total width. Though perhaps unlikely from a theoretical standpoint, it is possible that a larger branching ratio to dark matter could be compensated by larger couplings for the production of the Higgs, leaving the rates for the observed channels unchanged~\cite{Belanger:2013kya}.

The second method of measuring the Higgs width relaxes the requirement that the couplings to the fermions and gauge bosons are as in the Standard Model, and places a bound on the width via the measured interference of the Higgs and the $Z$. This constrains the Higgs width to be $\Gamma_h < 17.4$~MeV \cite{CMS-PAS-HIG-14-002,Khachatryan:2014iha}. However, as with the invisible Higgs decay measurement, this interference effect does make some assumptions about the production mechanism of the Higgs \cite{Englert:2014aca,hjets}. The third method remains fully agnostic as to the Higgs couplings. This is the most robust, but least constraining measurement: the direct measurement from the $h \to ZZ^* \to 4\ell$ channel, which has measured $\Gamma_h < 3.4$~GeV \cite{Chatrchyan:2013mxa}.

In Figure~\ref{fig:higgs_bounds}, we show the collider and direct detection constraints on the 125~GeV Higgs boson as a function of total width, assuming a coupling to dark matter $g_\chi$ (unlike Figures~\ref{fig:scalar_bounds} and \ref{fig:pseudo_bounds}, note that the horizontal axis is $\Gamma_h$, not $\Gamma_{\phi(A)}/m_{\phi(A)}$). As before, we parametrize the coupling to the Standard Model fermions as $g_v$. Given the present concordance between experiment and theory, the primary model-building focus for Higgs physics appears to be concentrating on scenarios with $g_v\sim 1$, and it appears to be difficult to find models where large deviations from the Standard Model prediction is consistent with all Higgs data in a realistic extension of the Standard Model~\cite{sfitter}. As we saw in the general scalar mediator, the collider bounds are much less constraining than those set by direct detection experiments. While the collider constraints are relatively insensitive to dark matter masses below $m_h/2$, the direct detection bounds weaken significantly significantly if the dark matter is below $\sim 6$~GeV.

It is surprising to see that the associated top channel is comparable here to the monojets, given the experimental difficulties in probing Standard Model $tth$ production. However, recall that the Standard Model search in this channel is forced to rely on $h\to b\bar{b}$ decay. If we assume a significant branching ratio of the 125~GeV Higgs into invisible dark matter, the much lower backgrounds in the dileptonic top plus MET channel allow the experiments to set a bound comparable to that of the monojets.

\section{Conclusions}

The next few years of data from the LHC Run-II has the potential to shed new light on the nature of dark matter. The EFT formalism has been very useful in the analysis of Tevatron and LHC data, allowing straightforward comparisons to direct and indirect searches, and moving dark matter searches in a more model-independent direction. However, the powerful bounds set by the LHC push the theory into a regime where the EFT often does not generally apply. This should be a cause for optimism: the break-down of the consistency of the EFT implies that, for much of the parameter space, if the LHC can produce dark matter then it can also produce associated particles that mediate the interaction between the dark sector and our own.

Previous works have introduced various simplified models which bridge the theoretical divide between the EFT and complete models such as supersymmetry. We add to this work by constructing two benchmark models of spin-0 mediators coupling to dark matter consisting of Dirac fermions. While such attractive models have been considered in the past, we -- for the first time -- provide a comprehensive set of constraints from direct detection, indirect searches, and three collider channels associated with missing transverse energy. 

As previous works have noted, care must be taken when simulating scalar mediated missing energy searches at the LHC, as these are primarily produced through a top-loop induced coupling to gluons. As the transverse momentum flowing through this loop is large compared to $2m_t$ (and may be large compared to the mediator mass), it has been demonstrated that working in approximations of infinite top mass and/or on-shell gluons can incorrectly predict the MET and $p_T$ distributions. In this paper, we clearly show the impact of these effects on the distribution of jet $p_T$ and MET, which are critical to missing energy searches at the LHC, and outline appropriate techniques for simulating these models. These issues will become even more important in future LHC runs, where higher energies will force harsher MET and $p_T$ cuts, further increasing the deviation between the distributions predicted by an effective operator treatment of the loop-coupling, and the correct one.

For our benchmark models, the monojet channel remains the most constraining out of all the collider bounds. However, associated heavy flavor searches are important; associated production with tops can rival the monojet channel in the low mediator mass region. As such these additional searches should be pursued as complimentary to the monojet bounds, sensitive to different combination of couplings. Similarly, the direct detection bounds place much more powerful limits on the couplings for scalar models, assuming the dark matter mass is heavier than $\sim 6$~GeV. However, there are astrophysical uncertainties inherent to direct detection limits, and the LHC searches provide an complimentary testing ground, one that independent of the uncertainties on our local dark matter density and velocity distribution. Similar astrophysical uncertainties also relate to bounds placed by indirect detection, and further collider searches may be a key factor in resolving the active debate about claimed signals from the Galactic Center. As can be seen from Figure~\ref{fig:pseudo_bounds}, the current constraints already touch on the relevant parameter space for $m_\chi \sim 40-50$~GeV, and can indeed rule out simplified models with mediators much heavier than 100~GeV as the source of the anomaly. Though modifications of the benchmark simplified model can explain the Galactic Center excess with particles that have vanishing LHC cross sections \cite{Abdullah:2014lla,Cheung:2014lqa}, it is interesting that one of the simplest scenarios is not yet ruled out, yet lies within realistic reach of the LHC in the near future. 

The searches we extracted bounds from in this paper were pre-existing and easily adapted to our simplified models. However, as should be clear, many other possible channels exist, which would place complimentary bounds on the couplings of our benchmark models. In addition to further missing energy searches in association with heavy flavor -- in particular, searches with $\tau$ leptons, which would probe the mediator-lepton coupling --  we suggest that future work should also consider the constraints from decays of mediators back into Standard Model particles. 

Given couplings $g_\chi$ and $g_v$ which are of the same order of magnitude, one would expect decays to dark matter to dominate. However, it is possible that $g_v \gg g_\chi$, or that the dark matter itself is kinematically inaccessible as a decay product of the mediator. In this second case, though some missing energy constraints can be placed from dark matter production via off-shell mediators, the collider production cross section of the mediator itself would be far higher. Channels with decays to $b\bar{b}$, $\tau \tau$, top pairs, or the experimentally clean $\gamma\gamma$ signatures are all likely candidates for dark matter simplified models, particularly with the spin-0 mediators considered here. If the width is small, than long-lived mediators are possible, and searches for displaced decays back to visible particles could place important limits on models with small $g_v$ and $g_\chi$ which would be otherwise inaccessible. CMS, ATLAS, and Tevatron have performed searches in some of the prompt channels, though their results are typically presented in terms of two-Higgs doublet models. As we have described in this paper, such signatures can be relevant to a large range of models, and could be an important part of our search for the new physics of dark matter.  

\section*{Acknowledgements}
MRB thanks Tim Tait, Dan Hooper, and Maria Spiropulu for helpful comments and suggestions.

\end{document}